 \documentclass[journal]{IEEEtran}
 \newcommand{\DRAFT}[1]{}
 \newcommand{\FINAL}[1]{#1}

\usepackage[noadjust]{cite}
\usepackage{graphicx}
\usepackage{amsmath}
\usepackage{xspace,bbm,amssymb}

\usepackage[latin1]{inputenc}
\usepackage{theorem}

\input alphabet
\input jidef

\def\dd{\,d}
\def\leq{\le}
\def\geq{\ge}
\renewcommand{\det}[1]{\bars{#1}}

\def\eqname{Eq.~}
\def\eqnames{Eqs~}
\def\tabname{Table~}

\def\figname{Figure~}

\def\chapname{Chapter~}

\def\parname{\S~}
\def\parnames{\S~}
\def\sectname{Section~}

\def\<{``}\def\>{''}

\def\SI{\text{if\:}}

\def\SINON{\text{otherwise}}

\def\barr{\begin{array}}                \def\earr{\end{array}}
\def\bcc{\begin{center}}                \def\ecc{\end{center}}
\def\cl#1{\centerline{#1}}
\def\bdoc{\begin{document}}             \def\edoc{\end{document}}
\def\ben{\begin{enumerate}}             \def\een{\end{enumerate}}
\def\beqn{\begin{eqnarray}}             \def\eeqn{\end{eqnarray}}
\def\beqnl#1{\beqn\label{#1}}           \def\eeqnl#1{\label{#1}\eeqn}
\def\beqnx{\begin{eqnarray*}}           \def\eeqnx{\end{eqnarray*}}
\def\bseqn{\begin{subeqnarray}}         \def\eseqn{\end{subeqnarray}}
\def\beq#1\eeq{\begin{equation}#1\end{equation}}
\def\bal#1\eal{\begin{align}#1\end{align}}
\def\balx#1\ealx{\begin{align*}#1\end{align*}}
\def\beqx{$$}                           \def\eeqx{$$}
\def\bfig{\protect\begin{figure}}       \def\efig{\protect\end{figure}}
\def\bfigx{\protect\begin{figure*}}     \def\efigx{\protect\end{figure*}}
\def\bfigt{\protect\begin{figurette}}   \def\efigt{\protect\end{figurette}}
\def\bfl{\begin{flushleft}}             \def\efl{\end{flushleft}}
\def\bfr{\begin{flushright}}            \def\efr{\end{flushright}}
\def\bit{\begin{itemize}}               \def\eit{\end{itemize}}
\def\bmi{\begin{minipage}}              \def\emi{\end{minipage}}
\newcommand{\bfmi}{\begin{fminipage}}   \newcommand{\efmi}{\end{fminipage}}
\def\bpic{\begin{picture}}              \def\epic{\end{picture}}
\def\btabl{\begin{table}}               \def\etabl{\end{table}}
\def\btab{\begin{tabular}} 
\def\btabu{\begin{tabular}}             \def\etabu{\end{tabular}}
\def\btabx{\begin{tabular*}}            \def\etabx{\end{tabular*}}

\def\qmod{\bars{q}}
\def\nupsf{\ensuremath{\nu_{\mathrm{PSF}}}}
\def\nuspec{\ensuremath{\nu_{\mathrm{spec}}}}
\def\nuharm{\ensuremath{\nu_{\mathrm{harm}}}}
\def\Dpsf{\ensuremath{\Dc_{\mathrm{PSF}}}}
\def\Dspec{\ensuremath{\Dc_{\mathrm{spec}}}}
\def\Cum{\mathrm{Cum}}
\renewcommand{\overline}[1]{{#1}^*}
\renewcommand{\Esp}[2][]{\mathcal{E}{_{#1}} \kern-2pt \left\{#2\right\} }
\renewcommand{\egdef}{:=}
\def\tr{\mathrm{Tr}}
\def\vectb{\textbf{vect}}

\newtheorem{prop}{Property}
\newtheorem{rmk}{Remark}

\usepackage{color}
\usepackage{soul}
\definecolor{violet}{rgb}{.56,0,1}
\definecolor{darkgreen}{rgb}{0,.6,0}
\definecolor{orange}{rgb}{1,.8,0}

\def\addJIF#1{{{#1}}}
\def\addJIU#1{{{#1}}}
\def\addJID#1{{{#1}}}
\def\addMAU#1{{{#1}}}
\def\addMAF#1{{{#1}}}
\def\addMAD#1{{{#1}}}
\def\addMAT#1{{{#1}}}

\def\add#1{{#1}}

\begin{document}

\title{
On the super-resolution capacity of imagers using unknown speckle illuminations
}
\author{\textbf{\today} -
 J\'er\^ome~Idier,~\IEEEmembership{Member,~IEEE, }
 Simon~Labouesse, \IEEEmembership{}
Marc~Allain, \IEEEmembership{}
Penghuan~Liu, \IEEEmembership{}
 S\'ebastien~Bourguignon,~\IEEEmembership{}
 and~Anne~Sentenac,~\IEEEmembership{Member, IEEE}
\thanks{The authors acknowledge partial financial support for this
  paper from the GdR 720 ISIS, the Agence Nationale de la Recherche
  under Grant ANR-12-BS03-0006 and the ITMO Cancer of the `Plan Cancer 2014-2019'.}
\thanks{J.~Idier, P.~Liu and S.~Bourguignon are with the CNRS and Ecole Centrale Nantes at the Laboratoire des Sciences du Num\'erique de Nantes (LS2N, CNRS UMR 6004), 
\DRAFT{1 rue de la No\"e, BP 92101,}%
F-44321 Nantes, France.
\DRAFT{Tel: (+33)-2 40 37 69 09, Fax: (+33)-2 40 37 69 30.}%
E-mail: firstname.name@ls2n.fr.}
\thanks{S. Labouesse, M. Allain and A. Sentenac are with Aix Marseille Univ, CNRS, Centrale Marseille, Institut Fresnel, Marseille, France. E-mail: firstname.name@fresnel.fr.}}

\maketitle
\begin{abstract}
Speckle based imaging consists of forming a super-resolved reconstruction of an unknown sample from low-resolution images obtained under random inhomogeneous illuminations (speckles). In a blind context where the illuminations are unknown, we study the \textit{intrinsic} capacity \addJIF{of speckle-based imagers} to recover spatial frequencies \addJIF{outside the frequency support of the data,}
\addJID{with minimal assumptions about the sample.}
We demonstrate that, under physically realistic conditions, the \addJIF{covariance} of the data has a super-resolution power corresponding to the squared magnitude of the imager point spread function. This theoretical result is important for many practical imaging systems such as acoustic and electromagnetic tomographs, fluorescence and photoacoustic microscopes, or synthetic aperture radar imaging. A numerical validation is presented in the case of fluorescence microscopy.
\end{abstract}

\section{Introduction}

In most active wave imaging systems, the recorded data $z$ can be modeled as the convolution of a point spread function
(PSF) $h$ with the product of the sample $\rho$ with an illumination $E$, plus some additive noise $\varepsilon$:
\beq
\label{eq:observation0}
z = h\otimes(\rho E) + \varepsilon
\eeq
where $\otimes$ stands for the convolution operator, either in two or three spatial dimensions. This simple model applies to imaging configurations as diverse as microwave scanners or anechoic chambers~\cite{VandenBerg03}, radar remote sensing~\cite{Munson83} or fluorescence microscopy~\cite{Gustafsson08}.

The \addJIF{shape of the point spread function $h$ depends on 
the imager geometry, \eg} the numerical aperture (NA) of the microscope
objective, or the size of the antenna array in radar imaging. It
accounts for the wave propagation from the sample to the detector. In
most configurations, free-space propagation prevents the wavefield 
high frequencies from reaching the detector. As a result, $h$ has
necessarily a bounded Fourier support \Dpsf. For instance, in a 
microwave scanner, \Dpsf\ is a \addMAD{hollow sphere} of radius $1/\lambda$ (where
$\lambda$ is the illumination wavelength) when the field scattered by the sample is recorded under all possible directions, or a cap of
sphere when the observation is performed only over a small solid angle \cite{Haeberle10}. Similarly, in two (or three) dimensional
fluorescence microscopy, \Dpsf\ is a disk (or a solid torus) of radius $\addJID{2}\text{NA}/\lambda$~\cite{Gustafsson08}.

When the illumination is homogeneous throughout the target, solely the
sample frequency components in \Dpsf\ can be restored from the data by linear methods,
which limits fundamentally the image resolution. To improve the
latter, synthetic imaging using multiple illuminations has been
developed. Its main principle is to use several known inhomogeneous 
illuminations $E_m, m=1\ldotsv M$, to probe the sample. 
The frequency mixing of $E_m$ with $\rho$ causes a down-modulation
of the sample high spatial frequencies into the frequency-support
\Dpsf
. Using appropriate
data processing, sample frequencies beyond \Dpsf\ can be recovered, 
yielding a much better resolution. This idea is at the core of many 
imaging configurations such as Synthetic Aperture Radar 
(SAR)~\cite{Munson83}, diffraction tomography~\cite{Devaney82}, 
and Structured Illumination fluorescence Microscopy (SIM) 
\cite{Heintzmann99,Gustafsson00}, among others.

In all these imaging modalities, the standard numerical or analog
process that forms the super-resolved image from the stack of 
low resolution data assumes the precise 
knowledge, and thus the tight control, of the different illuminations
$E_m$. The super-resolution capacity of the process is then both 
theoretically and practically demonstrated. However, the full control 
of the illumination patterns is a major constraint for the
experimental implementation and in some cases proves impossible. 
\addJIU{%
The case of thick samples imaged with three-dimensional SIM is a
classical example since samples are likely to introduce distorsions 
on the excitation pattern~\cite{Jost15,Ayuk13}.}
Hence, some groups have developed reconstruction algorithms able 
to handle some uncertainty about the illuminations~\cite{Thibault09,Wicker13,Ayuk13,Jost15}. Following a 
\addJIF{less conventional} option, others have advocated using of totally
uncontrolled illuminations of speckle type~\cite{Goodman07}. 
\addJIU{This recent blind approach could dramatically simplify the
  experimentation by further relaxing the constraints on controlling
  the illumination patterns. Examples of implementations can be found}
in optical microscopy~\cite{Mudry12,Min13,Oh13,Negash16,Labouesse17} and
photoacoustic imaging~\cite{Gateau13,Chaigne16}.
%
%
\addJIF{The proposed inversion schemes take advantage of 
the nonnegativity of the sample $\rho$, and on statistical information 
on $\rho$ and/or the illuminations $E_m$. In particular, some of them 
introduce sparsity information on $\rho$~\cite{Min13}, or on the 
products $\rho E_m$~\cite{Labouesse17}.}
Generally speaking, the stack of low resolution speckle data yielded reconstructed images with significantly better resolution than that provided by a standard imager using homogeneous illumination. 
However, many questions remain unanswered about the theoretical resolution that one can expect from such a system, in particular with respect to the speckle statistics. 

To the best of our knowledge, this paper provides the first comprehensive mathematical understanding of the super-resolution (SR)
capacity of synthetic imaging using speckle illuminations in a blind way. Our analysis is very general and basically holds when the data can be modeled by Eq. \eqref{eq:observation0} with $\rho\geq 0$. 
\add{%
Pivotal SR results are for instance provided for two popular
microscopy modalities, namely optical \textit{fluorescence 
microscopy} and optical \addMAD{\textit{coherent imaging}.}
} 

\add{\indent%
Fluorescence microscopy is an incoherent imaging modality, for which the 
quantities involved in \eqref{eq:observation0} have the following physical interpretation:
$\rho$ is the fluorescence density distribution, \ie a real-valued, nonnegative, quantity;
the incoherent PSF $h$ is real-valued, nonnegative, and \add{the support of} its Fourier transform is \add{a disk or a torus in two and three dimensions, respectively} \cite{Gustafsson08};
the illumination $E$ is a speckle intensity pattern produced by a coherent light beam, \ie a random real-valued, nonnegative,
quantity (see for instance \cite{Mudry12});
and $z$ is an intensity measurement plagued by real-valued instrumental noise $\varepsilon$. 
In addition, in the low counting-rate regime, one may consider photon 
counting fluctuations in the observation model: in this case, the quantity
$h\otimes(\rho E)$ in \eqref{eq:observation0} is connected
to the mean of the counting statistics. The case of intensity
measurement plagued by both photon counting fluctuations 
and electronic noise is specifically addressed in Appendix~\ref{Poisson}.
}

\add{\indent%
\addMAD{In coherent imaging}, such as tomographic diffraction
  microscopy \cite{Haeberle10}, 
we have the following correspondence for the model \eqref{eq:observation0}:
$\rho$ is the  relative permittivity contrast distribution, a
complex-valued function in general, although our mathematical analysis 
is restricted to real nonnegative $\rho$, 
\ie it is restricted to lossless dielectric objects; 
the coherent PSF $h$ is complex-valued, and \add{the 
support of} its Fourier transform is a sphere cap \cite{Haeberle10};
the illumination $E$ is a complex-valued random field, \eg a
circular Gaussian random field if it arises from 
\addMAD{a (scalar) electric field stemming from a fully developed speckle produced by coherent light 
\cite{Goodman85}};
\addMAD{the recorded data $z$ is the scattered 
electric field}  plagued\footnote{%
\add{This measurement $z$ is usually obtained by the Fourier transform 
of measured real valued intensities in an off-axis interferometric
mounting, see for instance \cite{Cuche99,Godavarthi15}. Moreover, if 
the counting rate is high enough, an additive  fluctuation model over 
the real and the imaginary part is relevant.}} by circular
complex-valued instrumental noise $\varepsilon$, such that 
$z$ is also a circular complex-valued random field. 
}

\addMAT{%
In the sequel, the term \textit{super-resolution} is understood 
as the ability to recover spatial frequencies of the sample that
cannot be obtained with either a constant illumination 
in incoherent imaging, or a single plane-wave with normal incidence 
in coherent imaging.      
}%
\addMAF{%
Following standard results \cite[Chap. 6]{Goodman96}, we recall that 
the incoherent (intensity) PSF is obtained by squaring the magnitude of the
coherent (complex electric field) PSF of the optical system. As a
result, transmitted spatial frequencies with an incoherent illumination 
span twice the domain transmitted with a coherent illumination. 
However, incoherent illuminations do not provide the permittivity 
contrast, but its squared magnitude, which prevents any direct 
comparison in terms of spatial resolution between coherent and 
incoherent optical systems, see \textit{e.g.,} \cite[Sec. 6.5]{Goodman96}  
for details. In contrast, the present work shows that a double spatial resolution can be obtained in \textit{both}
cases thanks to random illuminations.
}

\add{\indent%
We finally note that the model \eqref{eq:observation0} also encompasses some other situations,
namely \textit{microwave imaging} \cite{Jofre90} (the measured data are complex
fields, $\rho$ is the complex permittivity; the noise being mainly
an electronic fluctuation, it can be assumed Gaussian for both the real and 
imaginary parts) and \textit{photo-acoustic imaging}
\cite{Gateau13,Chaigne16}  (the measured data are real-valued 
B-mode images corrupted by real-valued Gaussian noise, and $\rho$ 
represents the optical absorption). Hereafter, we consider
the complex-valued setting, since the real setting can be deduced 
straightforwardly as a particular case where the imaginary parts of 
the relevant quantities vanish.
}



\add{\indent%
The article is organized as follows. The next section provides
the mathematic assumptions required in our SR analysis. 
Section~\ref{moments} establishes the expression of the first two
moments of the data. In section~\ref{sec:SR}, the dependency 
between the latter expressions and the spatial frequency components 
of the sample $\rho$ is further examined. Clear conclusions about the
SR capacity of the system are obtained if the speckle illuminations
are ``sufficiently'' correlated, in the sense that 
\addMAD{their spectral power density lies within the}
frequency support of the PSF. \add{Such conclusions constitute the main contribution of this paper}. The opposite case of 
uncorrelated speckles is also considered. Section \ref{sec:numerical}
deals with the practical question of a computational scheme to 
reconstruct the unknown scene. A two-dimensional simulation 
of an optical fluorescence microscope using  correlated speckle 
illuminations  is provided, and it supports that the expected SR 
ratio can be obtained from the data empirical second-order 
statistics.
}%
Finally, Section~\ref{sec:appli} discusses the practical consequences
of the obtained results, and evokes possible extensions and remaining points to address.

\section{Model and assumptions for the SR analysis}
\label{sec:assumptions}

We consider $M$ images $(\zb_1\ldotsv\zb_M)$ of the same sample that
have been acquired using $M$ different speckle illuminations. Each
image $\zb_m=(z_m(\rb_1)\ldotsv z_m(\rb_N))$ is a set of $N$ pixels,
each of which being indexed by a spatial coordinate vector $\rb_n$. In
practice, vector $\rb_n$ spans a finite $d$-dimensional rectangular
grid \Gc, common to all images, $d$ being equal to two or three. 
\add{Without loss of generality, we  consider the spatial sampling rate to be normalized to unity in each direction.}
\add{%
By convention, we  also consider that $\zb_m$ are column vectors obtained by scanning the image grid \Gc in an arbitrary order. Hence,
for all $m \in \{1 \cdots M\}$ and $\rb\in\Gc$, the observation model reads
\beq
 z_m(\rb) = y_m(\rb) + \varepsilon_m(\rb), 
 \label{eq:observation}
\eeq
with 
\bal
\label{eq:observationbis}
y_m(\rb) 
& = \int h(\rb-\rb') \rho(\rb') E_m(\rb')\dd \rb',
\eal
and where $E_m$ and $\varepsilon_m$ are random quantities: $E_m$ is the $m$th random illumination and $\varepsilon_m$ stands for electronic noise. 
}
Furthermore, the following assumptions will be made concerning the observation model \eqref{eq:observation}: 
\bit
\item[($i$)] 
The PSF $h$ is both integrable and square-integrable
 (\ie $\int \bars{h(\xb)}^p\dd\xb<\infty$ for $p=1,2$). 
\addMAD{%
Moreover, its Fourier transform\footnote{%
Hereafter, the tilde sign $\,\wt{}\,$ denotes the $d$-dimensional continuous-space Fourier transform.
} $\wt{h}$ takes finite values and vanishes outside 
a bounded set $\Dpsf=\stdacc{\ub ~|~ \wt{h}(\ub) \neq 0}$.
}
%
%
These assumptions are met when the measurement 
$\zb_m$ is obtained in the far-field domain, which is 
the case of most imaging systems~\cite{Haeberle10}.
\item[\addJIF{($ii$)}] The sample $\rho$ is integrable and takes finite, nonnegative values over $\eR^d$. \addJIF{Moreover, it approaches zero at large distance from the origin}.
%
\item[\addJIF{($iii$)}] \addJIF{The data grid \Gc is sufficiently
    large to make the influence of finite data extent negligible. As a
    consequence, we will identify $\Gc$ with $\eZ^d$ in the sequel. 
    This is indeed  a legitimate simplification given Assumptions
    ($i$) and ($ii$) since we can show that $h\otimes(\rho E)$ tends 
    towards zero at large distance from the origin, provided that the 
illumination pattern $E$ is bounded.}
%
%
\item[\addJIF{($iv$)}] \add{$\Gc$ is fine enough to sample the PSF $h$ with no discretization error. According to Parzen's multidimensional extension of Shannon theorem \cite{Parzen56}, such a condition is met as soon as $\Dpsf$ belongs to the baseband $\Bc=[-1/2,1/2]^d$.}
\item[\addJIF{($v$)}] The noise and illuminations are second-order stationary,
mutually decorrelated random processes. This is a standard
hypothesis which is verified for most imagers
\cite[Sec.~4.4]{Goodman07}. Moreover,  a direct extension would be
possible  to cases where the statistical mean of the illuminations is 
spatially varying. Without loss of generality, we will also assume
that the noise is zero-mean.
\item[\addJIF{($vi$)}] The first two moments of the illuminations \addJID{and of the noise} are known. This assumption is at the core of our approach. It is expected to be less difficult to satisfy than the knowledge of the illuminations patterns.
\eit



In this work, we restrict the analysis of the data by considering only 
second-order statistics, \ie the statistical mean and covariance of
the data. More precisely, our aim is to determine the spatial
frequency domain over which the sample spectrum can be identified 
from these statistics. Such a restriction is legitimate for several reasons.

On the one hand, the empirical mean and covariance are easily
accessible statistical quantities, that can provide reliable
second-order information from a practically acceptable number of illuminations. 

On the other hand, the statistical mean and covariance are exhaustive
statistics if the data are Gaussian, whether it is real-valued or
complex circular. For instance, the latter assumption is suited to 
coherent imaging techniques such as \addMAD{tomographic diffraction  microscopy}.
%
In other situations, such as optical fluorescence microscopy, the speckle illumination, and hence
the data, are non-Gaussian. 
The statistical mean and covariance do not summarize all the information about the sample available in the measurements in such situations, but our results still provide a \textit{lower bound} on the information retrievable from the complete data statistics.

With the goal of characterizing the SR potential of second-order methods, we wish to assign each component of the spatial Fourier transform $\wt{\rho}(\ub)$ of the imaged sample to one of the three classes:
\bit
\item \textit{Non-identifiable} spectral components are those for which the second-order data statistics bring no information.
\item \textit{Partially identifiable} components are those for which the second-order data statistics bring some information, but for which some ambiguity remains.
\item \textit{Identifiable} components are those which are uniquely determined given the second-order data statistics.
\eit
Obviously, the \add{support of each class in the Fourier domain} may depend on the frequency support \Dpsf\ and on the \addJIF{covariance} structure of the speckle illumination.

\section{First and second-order statistics of the data}
\label{moments}

The statistical mean and \addJIF{covariance} of the data are now derived. In what follows, \add{$\Esp{\cdot}$ and $^*$ denote the statistical 
expectation operator and complex conjugation, respectively}. According to \addJIF{assumptions ($v$) and ($vi$)}, let $\Esp{E}=E_0$ 
and $\gamma_E(\rb)=\Esp{E(\xb)\overline{E}(\xb-\rb)}-\bars{E_0}^2$ denote the mean and \addJIF{covariance} function of the speckle, and 
let $\gamma_\varepsilon(\rb)=\Esp{\varepsilon(\xb)\overline\varepsilon(\xb-\rb)}$ denote the \addJIF{covariance} function of the noise.

\subsection{First-order information content}

From the observation model \eqref{eq:observation}-\eqref{eq:observationbis} and from the assumption of centered noise, we deduce the statistical mean:
\beq
 \label{eq:mu_z}
\mu_z(\rb) =
\Esp{z_m(\rb)} 
= E_0\int h(\rb-\rb') \rho(\rb')\dd \rb',
\add{\quad \rb\in\eZ^d.}
\eeq
The continuous-space Fourier transform of $\mu_z$ reads
\beq
\wt\mu_z(\ub)=E_0\,\wt h(\ub) \wt{\rho}(\ub)
\label{TFmu}
\eeq
for all $\ub\in\eR^d$.
Function $\wt\mu_z$ has a support limited to $\Dpsf$, so, according to assumption \addJIF{($iv$)}, the sampling of $\mu_z$ on $\eZ^d$ is lossless.
%
%
A straightforward deduction from expression \eqref{TFmu} is that any spectral component of $\rho$ belonging to the support \Dpsf\ is identifiable, provided that $E_0\neq 0$. In particular, if $E_m(\rb)$ is a complex circular Gaussian process, then $E_0=0$~\cite{Goodman07} and $\mu_z(\rb)$ brings no information about the unknown sample. This conclusion leads to the following property.
\begin{prop}
 \label{prop_D1}
  The frequency component $\wt{\rho}(\ub)$ is identifiable from $\mu_z$ if and only if $\ub\in \Dc_1$ with 
 \begin{equation}
 \label{Def_D1}
 \Dc_1 = \left\{
  \begin{array}{ll}
  \nonumber
  \Dpsf\ &\quad \text{if~} E_0 \neq 0,\\
  \varnothing &\quad \text{otherwise}.
  \end{array}
 \right.
 \end{equation}
\end{prop}
In any case, the first-order moment does not convey any information on the spectral components outside \Dpsf, \ie it brings no SR capacity. 

\subsection{Second-order information content}
Now let us focus on the data \addJIF{covariance} function \add{that reads}
\beqx
\gamma_z (\rb,\rb') = \Esp{z_m(\rb) \overline{z}_m(\rb')} - \mu_z(\rb)\,\overline{\mu}_z(\rb')
\eeqx
\add{with $\rb$, $\rb'\in \eZ^d$. 
If \eqref{eq:observation} and \eqref{eq:observationbis} hold, then we immediately get
\bal
\gamma_z(\rb,\rb') & =\gamma_y(\rb,\rb')+\gamma_\varepsilon(\rb-\rb')
\label{eq:correlation_z}
\eal
with
\bal
\gamma_y&(\rb,\rb') =\notag\\
&\iint\rho(\xb)\rho(\xb')\,h(\rb-\xb) \overline{h}(\rb'-\xb')\,\gamma_E(\xb-\xb') \dd\xb \dd\xb'.
\label{eq:correlation_y}
\eal
%
}
\addJID{The noise covariance function $\gamma_\varepsilon$ is 
known according to assumption ($vi$), but it conveys no information 
about the sample. {The knowledge of $\gamma_z$ is thus equivalent 
to that of $\gamma_y$, the latter term being the only potential source 
of information about the spectral components outside $\Dpsf$}}.
Let us examine the Fourier content of $\gamma_y$, first neglecting its discrete character. Using the continuous-space Fourier transform of 
\eqref{eq:correlation_y}, we obtain
\bal
\wt{\gamma}_y(\ub,\ub')
&=\addMAD{\wt{h}(\ub)\,\wt{h}^*(-\ub')}
\FINAL{\notag\\&}
\iint e^{-2i\pi(\ub\cdot\xb+\ub'\cdot\xb')}\rho(\xb)\rho(\xb')\gamma_E(\xb-\xb')\dd\xb \dd\xb'
\eal
for all $\ub$, $\ub'\in\eR^d$, where $\ub\cdot\rb$ denotes the usual scalar product in $\eR^d$.
Given that
$$
\gamma_E(\xb-\xb')=\int e^{2i\pi(\xb-\xb')\cdot\ub''}\wt{\gamma}_E(\ub'')\dd\ub'',
$$
it is easy to express $\wt{\gamma}_y(\ub,\ub')$ as follows:
\beq
\label{eq:tgammay}
\wt{\gamma}_y(\ub,\ub')=
\addMAD{\wt{h}(\ub)\,\wt{h}^*(-\ub')}\,
\wt{g}(\ub,\ub'),
\eeq
with
\beq
\label{eq:g}
\wt{g}(\ub,\ub')=\int \wt{\rho}(\ub-\ub'')\,\wt{\rho}(\ub'+\ub'')\,\wt{\gamma}_E(\ub'')\dd\ub''.
\eeq
According to \eqref{eq:tgammay} and to assumption \addJIF{($iv$)}, $\wt{\gamma}_y$ has a support limited to $\Bc\times\Bc$. 
Hence, $\wt{\gamma}_y(\ub,\ub')$ identifies with the discrete-space Fourier transform of $\gamma_y$ for all $\ub$, $\ub'\in\Bc$.
\addJIF{We conclude that the available information on the sample $\rho$ from the discrete data covariance is contained in (and limited to) $\wt{g}(\ub,-\ub')$, for all $\ub$, $\ub'\in\Dpsf$}.

\section{Super-resolution capacity of second-order methods}
\label{sec:SR}

\add{According to expressions \eqref{eq:correlation_y} or \eqref{eq:tgammay}-\eqref{eq:g},  the spectral density $\wt{\gamma}_E$  clearly plays a central role in identifying the spectral components of the sample. However, a difficulty in analyzing the SR capacity of second-order methods comes from the fact that 
the data \addJIF{covariance} is not a linear but a quadratic functional
of the unknown sample $\rho$. As a consequence, no general theory can
be applied to solve equations \eqref{eq:correlation_y} or
\eqref{eq:tgammay}-\eqref{eq:g} for $\rho$. However, two cases lend
themselves to a deeper analysis. 
The first one corresponds to ``sufficiently'' correlated speckles, 
in the sense that  
\addMAD{the frequency support of the covariance function $\wt\gamma_E$}
is contained in the frequency support of 
the PSF. At the opposite, the case of uncorrelated speckles can 
also be treated. These two cases are examined in the next two subsections,
whereas handling the intermediate case remains an open issue.}
\add{In the sequel, we make use of the Minkowski difference between two sets
$$
A\ominus B=\acc{\xb-\yb, \xb\in A,\yb\in B}
$$
to define the frequency domains over which the identification 
(or partial identification) of the frequency components of 
the sample is possible.
}

\subsection{Case of ``sufficiently'' correlated speckle}
\label{ssec:correl}

Let us assume that the unknown speckle illuminations are spatially
correlated, and \addJID{that the frequency support of its covariance
  function is $\Dspec=\{\ub ~|~\wt\gamma_E(\ub) \neq 0\}$.} 
According to expression \eqref{eq:tgammay},
$\wt{\gamma}_y(\ub,\ub')$ vanishes when either $\ub$ or $-\ub'$ is
outside \Dpsf. On the other hand, according to \eqref{eq:g},
$\wt{g}(\ub,\ub')$ conveys no information on the frequency components
$\wt{\rho}(\vb)$ such that either $\vb \pm \ub$ or $\vb \pm \ub'$
falls outside \Dspec. Then, the following property holds.

\begin{prop}
 \label{numax}
 Any spectral component $\wt{\rho}(\ub)$ such that $\ub \not\in\Dc_1\cup\Dc'_2$ with $\Dc'_2=\Dpsf\ominus\Dspec$ is non-identifiable from 
 the mean $\mu_z(\rb)$ and the \addJIF{covariance} function $\gamma_y(\rb,\rb')$.
\end{prop}

\begin{rmk}
 \label{rmk}
If each speckle pattern was known, the set of identifiable 
frequency components would be $\Dc_1 \cup \Dc'_2$ \addJID{for a 
sufficiently large number of speckles.}
Moreover, the components outside $\Dc_1 \cup \Dc_2'$ would remain
non-identifiable. 
In the same way, if the complete data statistics were available (the 
speckle patterns being unknown), the components outside $\Dc_1 \cup
\Dc_2'$ would also be non-identifiable, since the latter situation is
not more favorable than the former. We thus conclude that frequency 
components outside $\Dc_1 \cup \Dc'_2$ cannot be retrieved from 
standard (\ie non Bayesian) statistical information, even including higher moments.
\end{rmk}


Property~\ref{numax} is of negative nature. Fortunately, a positive
partial converse can be established in the important situation where 
the frequency support of the illuminations \addMAD{$\Dspec$} is not larger than that of 
the PSF. The following non trivial property holds. Its proof is reported in Appendix~\ref{proof}.
\begin{prop}
 \label{Mercer}
Provided that $\gamma_E$ is such that $\Dspec\subseteq\Dpsf$, any spectral component $\wt{\rho}(\ub)$ is identifiable from the mean $\mu_z(\rb)$ and the \addJIF{covariance} function $\gamma_y(\rb,\rb')$ if $\ub \in \Dc_1 \cup \Dc''_2$ with $\Dc''_2= \Dspec\ominus\Dspec$.
\end{prop}

\add{%
\begin{rmk}
\label{rmkSecIII_1}
An alternative definition of $\Dc''_2$ is obtained \textit{via} 
\begin{equation}
 \label{Def_D2bis}
 \Dc''_2 = \stdacc{\ub ~|~ (\wt{\gamma}_E\star\wt{\gamma}_E) (\ub) > 0}
\end{equation}
where $\star$ is the (deterministic) cross-correlation\footnote{%
\add{The cross-correlation between two square-integrable functions $f_1$ and $f_2$  is defined by
$$
(f_1 \star f_2)(\xb) = \int f_1^*(\xb') f_2(\xb' + \xb)\,\text{d}\xb'. 
$$
}
} operator. 
\end{rmk}
}

Let us consider a two-dimensional 
\addMAF{(2D) incoherent (\textit{e.g.,} fluorescence)} 
microscopy  problem as an illustrative example for Properties~\ref{prop_D1}, \ref{numax} 
and~\ref{Mercer}. In this case, \Dpsf\ and \Dspec\ are  centered disks 
of respective radii $\nupsf$ and $\nuspec$ and  $\Dc''_2$ is a centered disk 
of radius $2\nuspec$. As a consequence, if $\nupsf/2<\nuspec<\nupsf$,
we get $\Dc_1 \subset \Dc''_2$ and a SR factor of $2\nuspec/\nupsf$.

\begin{figure}[hbt]
\centering
\includegraphics[height=6.cm]{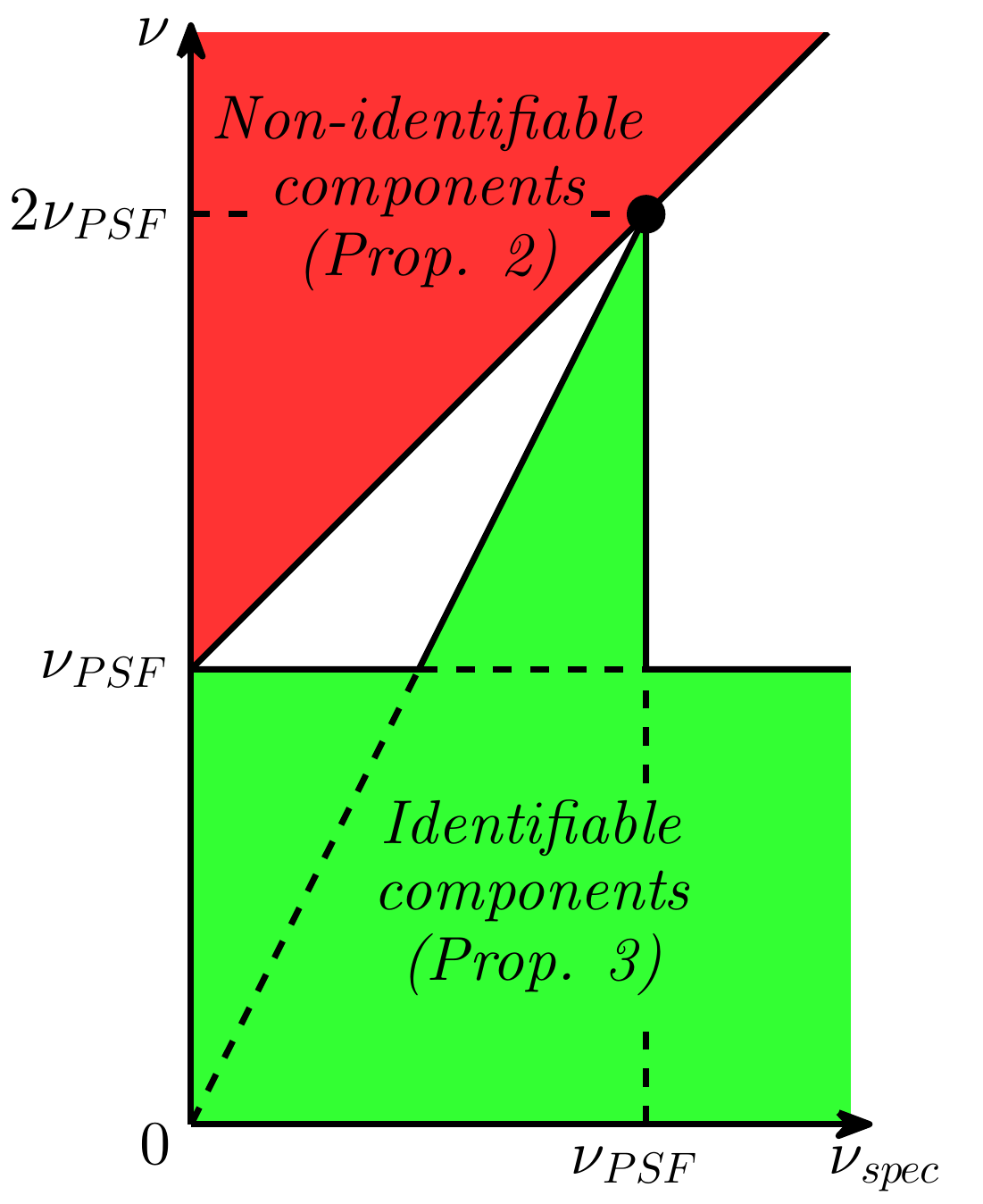}
\caption{%
Illustration of Props. \ref{numax} and \ref{Mercer} when \Dpsf\
 and \Dspec\ are centered disks of respective radii $\nupsf$ and
 $\nuspec$ (and $E_0\neq 0$). The cutoff frequency $\nupsf$ is fixed,
 while the speckle maximal frequency $\nuspec$ varies along
 the X-axis. The range of identifiable and of non-identifiable
 frequency components are represented along the Y-axis.
}
\label{fig:bsim}
\end{figure}

Figure~\ref{fig:bsim} gives a graphical illustration of the situation.
Let us remark that the status of the frequency components outside the
colored areas remains an open question. Our conjecture is that they
are only partially identifiable from the second-order data statistics.

\addJIU{\indent%
In the same conditions, according to Remark \ref{rmk}, the SR factor 
would be equal to $1+\nuspec/\nupsf$ if each speckle pattern was 
known. On Fig.\,\ref{fig:bsim}, such a limit corresponds to the 
boundary line of the set of non-identifiable components. In a similar way, 
classical (harmonic) SIM would yield an SR factor equal to $1+\nuharm/\nupsf$ 
if many known harmonic illuminations were used, at frequencies 
spread around a centered circle of radius $\nuharm$.}

\add{\indent%
The important case $\Dspec=\Dpsf$ is encountered in practice 
when illuminations and observations are performed \textit{via} the
same  components (same antenna array for emission and detection, or
same microscope objective for illumination and collection). The
latter  two properties then allow us to reach a tight conclusion in
this context:
}
second-order data statistics are sufficient to identify all the
frequency components  of the sample within 
$\Dc_2''\equiv\Dc_2'\equiv\Dc_2=\Dpsf\ominus\Dpsf$, and bring no
information  outside (such a situation corresponds to the black dot in 
Fig.~\ref{fig:bsim}). In other words, they should permit to recover
the sample  with a resolution equivalent to that of $|h|^2$, akin to
classical  SIM in fluorescence microscopy.

\begin{figure}[h]
 \centering
 \begin{tabular}{c}
  \includegraphics[width=6.4cm, bb=90 315 500 550, clip=true]{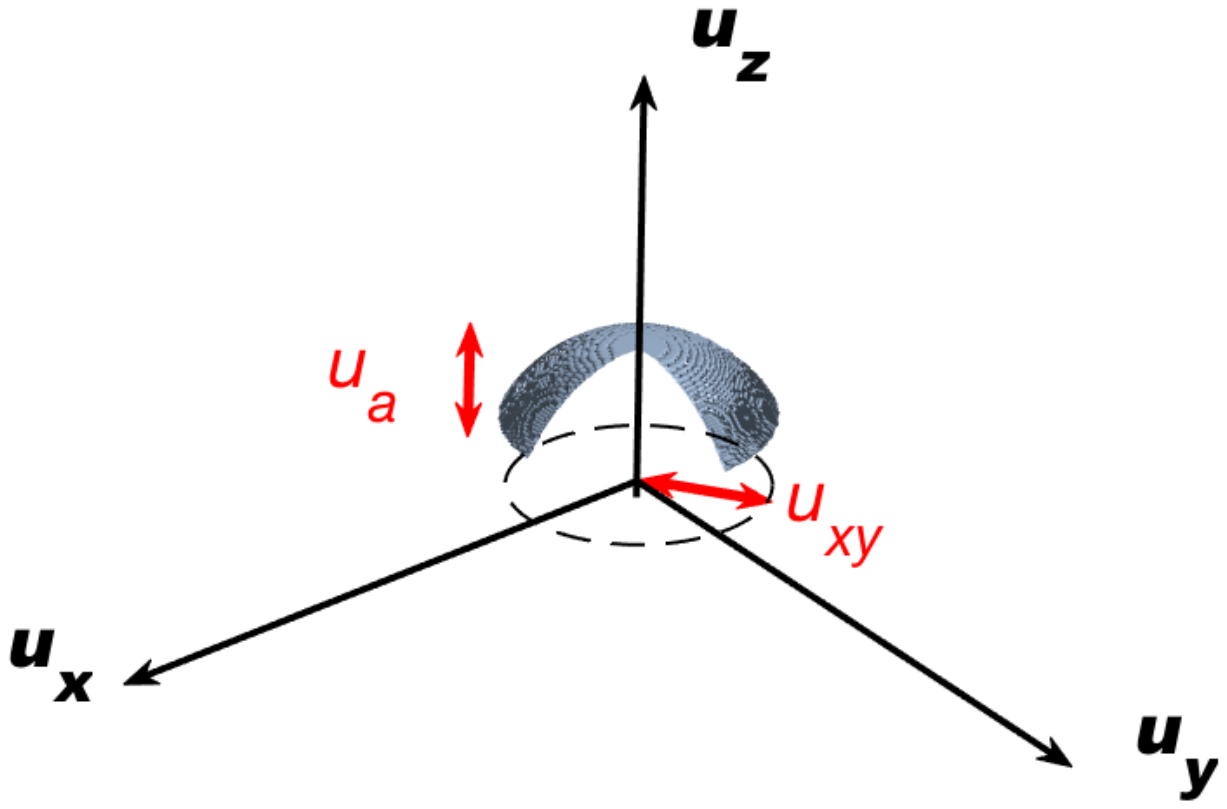}
 \\\small(a)\\[2em]
  \includegraphics[width=6.4cm, bb=90 290 520 520, clip=true]{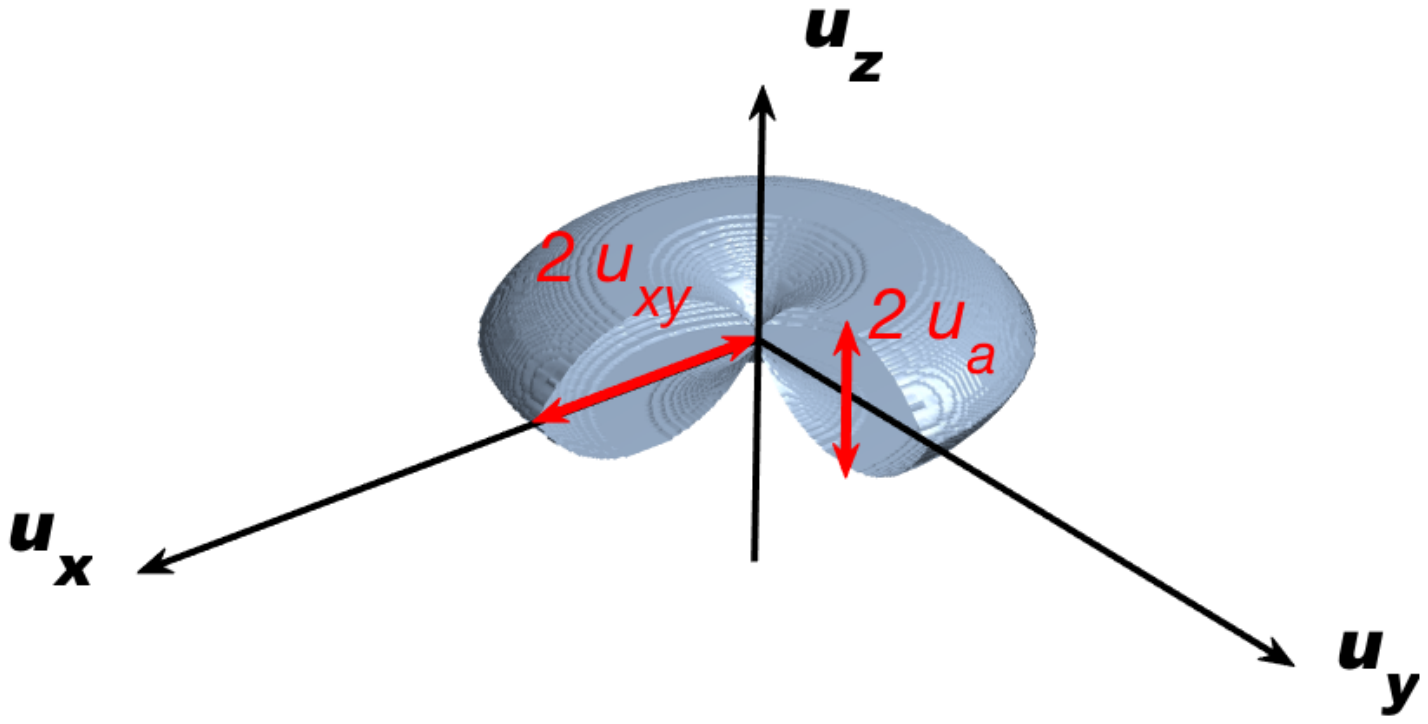}
 \\\small(b)\\[2em]
  \includegraphics[width=6.4cm, bb=90 290 520 520, clip=true]{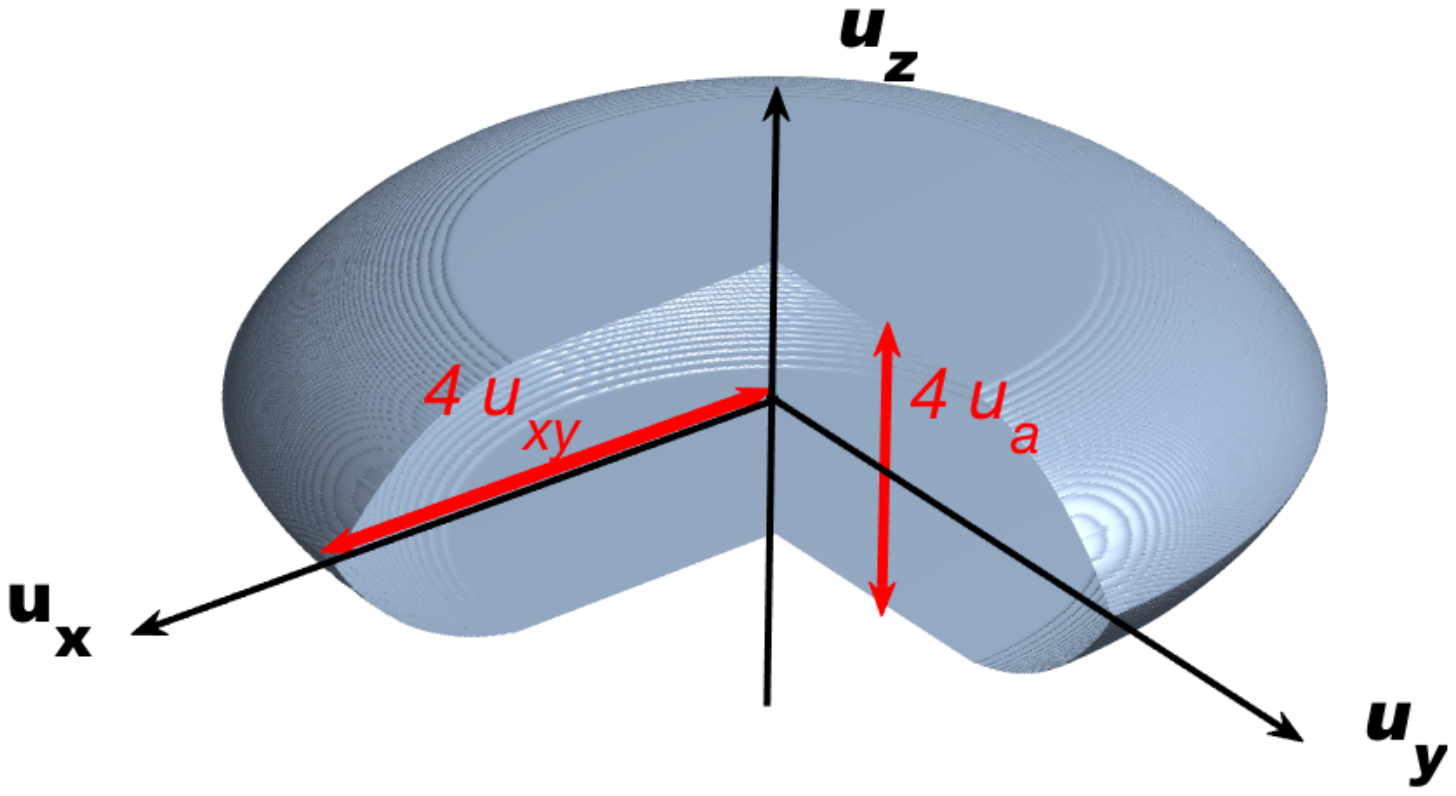}
\\\small(c)
\end{tabular}
 \caption{%
\addMAD{%
  {\bf Coherent imaging system}: (a) $\Dpsf$ is a surface in the 3D Fourier domain
with an isotrope \textit{lateral} cutoff frequency $u_{xy}$ and an
\textit{axial} frequency extension $u_a$;  (b) Assuming random excitations such that  
$\Dspec=\Dpsf$, the frequency components of the  permittivity 
contrast $\rho$ should be further identified over a domain 
$\Dc_2''$ that is a torus with a lateral (resp. axial) 
frequency extension of $2u_{xy}$ (resp. $2u_a$).
{\bf Incoherent imaging system}: (b) $\Dpsf$ is a solid torus 
exhibiting a ``missing cone'' along $u_z$; (c) Assuming random 
illuminations such that  $\Dspec=\Dpsf$, $\rho$ should be 
further identified over an extended frequency domain $\Dc_2''$
providing  a lateral (resp. axial) resolution of $4u_{xy}$ 
(resp. $4u_a$) without any ``missing-cone'' along $u_z$.
}
}
 \label{fig:dom}
\end{figure}

\addMAD{\indent%
Let us also add a few comments about our main result for 
three-dimensional (3D) problems:
\begin{itemize}
\item For \textit{coherent} imaging systems, $\wt h$ is typically a 
hollow spherical cap, as depicted in Fig.~\ref{fig:dom}(a), 
and thus a single incoming excitation (plane wave) $E$ cannot 
provide any 3D information about the permittivity contrast $\rho$ \cite{Haeberle10}.
In contrast, the same setup using coherent, but random excitations 
such that  $\Dspec=\Dpsf$, is able to retrieve 
$\wt\rho$ within a domain $\Dc_2''$  that is a centered solid 
torus, hence providing  3D information\footnote{%
\addMAT{%
The same identification domain can be obtained from a 
set of plane waves with various incoming angles, \textit{i.e.,}
with the additional  difficulty and slowness of controlling the 
angles of illuminations, see \cite{Haeberle10,Godavarthi15} for details.} 
}
about the permittivity contrast.
\item In \textit{incoherent} imaging, 
$\wt h$ is typically a solid torus, as depicted in Fig.~\ref{fig:dom}(b), 
which provides very poor sectioning capability along the axial direction $z$; this 
is the so-called ``missing-cone'' problem in wide-field 
incoherent microscopy \cite{Macias-Garza88}. In this case, 
speckle intensity illuminations such that $\Dspec=\Dpsf$ 
give access to a frequency domain $\Dc_2''$ that provides 
an extended lateral and axial resolution without any  
missing-cone, see Fig.~\ref{fig:dom}(c). 
This domain is actually equivalent to that of a 
perfect confocal microscope with an infinitively small pinhole 
\cite{Kimura90}, but it is obtained with no transverse 
scanning and no loss of photons.
\end{itemize}
}

\add{\indent%
As a final note, we stress that Property~\ref{Mercer} deals with the 
identifiability of the frequency components of the sample, and it does 
not predict the reachable estimation precision in the realistic
situation of a limited set of noisy data and of a limited field of view. Nevertheless, the numerical 
reconstruction example proposed in Section~\ref{sec:numerical}
suggests that \add{most of the} frequency components within $\Dpsf\ominus\Dpsf$ 
can be reliably retrieved in practice.}



\subsection{Case of uncorrelated speckle}
\label{ssec:uncorrel}
Let us now assume that $\gamma_E(\rb)=\gamma_E(\zerob)\delta(\rb)$, where $\delta$ is a Dirac distribution. This assumption can be considered valid when the speckle correlation typical size is much smaller than that of the point spread function. Then \eqref{eq:correlation_y} becomes
\beq
\label{eq:correlation_y_speckledec}
\gamma_y(\rb,\rb') = \gamma_E(\zerob)\int \rho^2(\xb) \, h(\rb-\xb) \overline{h}(\rb'-\xb) \,\dd\xb.
\eeq
In the Fourier domain, $\wt{\gamma}_E(\ub)=\gamma_E(\zerob)$, so 
\eqref{eq:tgammay} read
\bal
%
\label{eq:correlation_z_speckledec_TF}
\wt{\gamma}_y(\ub,\ub') &= \gamma_E(\zerob)
\addMAD{\wt{h}(\ub)\, \wt{h}^*(-\ub')} \, 
\wt{\rho^2} (\ub+\ub').
\eal
The latter relation is important since it yields that $\rho^2$ is accessible over all 
frequencies $\ub+\ub'$ such that both $\ub$ and $-\ub'$ belong to \Dpsf, \ie over the set 
$\Dpsf\ominus\Dpsf$. As a conclusion, the following property holds.
\begin{prop}
 \label{prop_decorr}
  The frequency component $\wt{\rho^2}(\ub)$ is identifiable from 
 the \addJIF{covariance} function $\gamma_y(\rb,\rb')$ if and only if $\ub\in \Dc_2 = \Dpsf\ominus\Dpsf$.
%
\end{prop}
A remarkable fact is that Property~\ref{prop_decorr} still holds if
only the variance $v(\rb)=\gamma_y(\rb,\rb)$ is considered instead of
the full \addJIF{covariance} function $\gamma_y(\rb,\rb')$\add{, provided 
that the data grid fulfills a more stringent condition than assumption \addJIF{($iv$)}. 
The starting point is the following relation obtained from \eqref{eq:correlation_y_speckledec}:
\beq
v(\rb) = \gamma_E(\zerob)\,(\rho^2 \otimes |h|^2)(\rb),
\label{eq:squared_conv}
\eeq
The continuous-space Fourier transform of $v$ reads
$$
\wt v(\ub)= \gamma_E(\zerob)\,\wt{\rho^2} (\ub) \times (\wt{h} \star \wt{h}) (\ub).
$$
Since the support of $\wt{h}\star\wt{h}$ spans the domain $\Dc_2$, 
the discrete-space Fourier transform of $v$ identifies with $\wt v$ provided that $\Dc_2\subset\Bc$. This assumption is stronger than $(ii)$. Typically, it means that the data must be acquired at least at twice the Nyquist rate. Alternatively, the acquisition rate could be unchanged, but then the data should be interpolated on a twice finer grid to yield the variance \eqref{eq:squared_conv}. \addJIF{Obviously, interpolation will not bring any new information. It will simply allow us to preserve the SR information contained in the sampled variance function $v$, and more precisely to avoid aliasing on $v$.} \addJID{On the other hand, interpolation will also apply to the noise component, so that the corresponding statistics should be modified accordingly.}
}
%
\begin{prop}
Provided that $\Dc_2\subset\Bc$, the frequency component $\wt{\rho^2}(\ub)$ is identifiable from the variance function $\gamma_y(\rb,\rb)$ if and only if $\ub\in \Dc_2$.
 \label{prop_decorr2}
\end{prop}
\add{A statement somewhat similar to Property \ref{prop_decorr2} was already made in~\cite{Oh13}, assuming a circular aperture and a 
  single incoherent point source (for $\rho$). Whereas the authors 
  of~\cite{Oh13} assert that uncorrelated speckle 
  illumination has the ability to improve the resolution of the 
  imaging setup beyond the diffraction limit, it should be stressed 
  that if  $\wt{\rho^2}=\wt\rho\otimes\wt\rho$ can be retrieved on
  $\Dc_2$, this does not mean that $\wt{\rho}$ can be
  retrieved on the same domain, nor in any other domain. 
  %
  }
\add{In practice, additional constraints (\eg the positivity of the
  sample assumed in this paper) can be further considered 
\cite[Sec. 6.6.4]{Goodman96}, \cite{Chaigne16}, but with 
 no formal guarantee about the super-resolution property 
obtained on $\rho$, to our best knowledge.
}

\section{Numerical implementation for 2D speckle SIM}
\label{sec:numerical}

The goal of the present section is to give a practical illustration of 
Property \ref{Mercer}, which is the main theoretical result of Section~\ref{sec:SR}. 
For this purpose, we consider a 2D speckle illumination fluorescence microscopy problem.
In the standard assumption of a perfect circular lens, $h$ is the
so-called Airy pattern~\cite[Sec. 4.4.2]{Goodman96}, and the optical 
transfer function (OTF) $\wt h$ defines a support $\Dpsf=\acc{\ub, 
\norm{\ub}_2< 2 \froc{\text{NA}}{\lambda}}$ with NA the numerical 
aperture of the microscope and $\lambda$ the emission/excitation 
wavelength. We further assume that the illumination of the sample 
and the collection of the emitted light is performed through the same 
optical device. Ignoring the Stokes-shift\footnote{%
The Stokes-shift \cite{Lakowicz06} implies that the support $\wt h$ is slightly smaller than the support of $\wt \gamma_E$. This difference between supports is 
small (about 10\%) and we expect that it will have a negligible impact on the SR effect that should reach twice the cutoff frequency of the OTF.}, we consider hereafter that $\gamma_E= E_0^2 \,h$. According to Property \ref{Mercer}, an SR effect approaching a factor two is expected from the empirical second-order statistics of a set of $M$ collected images, for asymptotically large values of $M$. The goal here is to show empirically that this SR effect does happen in realistic conditions, and in particular for moderately large values of $M$.

\subsection{Discretized model for 2D speckle SIM}

For the sake of computer implementation, \eqref{eq:observationbis} must be replaced by its discretized counterpart
\beq
\zb_m=\Hb\Rb\Eb_m+\varepsilonb_m,
\label{eq:observation3}
\eeq
where \Hb is a symmetric convolution matrix, and $\Rb=\Diag(\rhob)$, so that $\Rb\Eb_m$ corresponds to the product between the vectorized sample $\rhob$, and the vectorized $m$th illumination pattern $\Eb_m$. The mean vector and the covariance matrix of the acquisition $\zb_m$ now read
\beq
\label{mom_vect}
\mub_z=E_0\Hb\rhob,\qquad
\Gammab_z=\Hb\Rb\Gammab_E\Rb\Hb+\Gammab_\varepsilon,
\eeq
where $\Gammab_E$ and $\Gammab_\varepsilon$ are the covariance matrix of the speckle patterns and of the additive noise, respectively. For any finite number of illuminations $M$, the empirical mean $\hat{\mub}_z$ and covariance $\hat{\Gammab}_z$ are defined as
\beq
\wh{\mub}_z=\frac1M\sum_{m=1}^M\zb_m,\qquad\wh{\Gammab}_z=\frac1M\sum_{m=1}^M\zb_m\zb_m^\dag-\wh{\mub}_z\wh{\mub}_z^\dag,
\label{eq:empirical}
\eeq
where the symbol $^\dag$ stands for the transpose conjugate operator. As $M$ grows, $\hat{\mub}_z$ and $\hat{\Gammab}_z$ respectively converge toward $\mub_z$ and $\Gammab_z$, so that spatial frequency components of the sample within $\Dc_2''$ become identifiable, according to Prop.~\ref{Mercer}.
With a view to propose a computationally effective strategy to retrieve the latter components, we first directly formulate the main elements of Prop.~\ref{Mercer} and of its proof in a finite dimensional (\ie discretized) framework.

\subsection{Matrix transposition of Property \ref{Mercer}}
\label{sec:numtransp}

In the discrete framework of model \eqref{eq:observation3}, the matrix
formulation of Property~\ref{Mercer} mostly relies on the one-to-one
mapping between the asymptotically available covariance
$\Gammab_y=\Hb\Rb\Gammab_E\Rb\Hb$ and the matrix
$\Sb=\Gammab_E^{1/2}\Rb\Gammab_E^{1/2}$, provided that
$\Ker\Hb\subseteq\Ker\Gammab_E$. 
\addJIU{%
The latter condition is the discrete-space counterpart of the
condition $\Dspec \subseteq \Dpsf$ that allows the identification 
result stated in Prop.~\ref{Mercer}. ($\Ker\Mb$ denotes the set of 
vectors \vb such that $\Mb\vb$ is the null vector). Then,} we can 
show that \Sb is the unique Hermitian positive semi-definite\footnote{A 
Hermitian matrix is positive semi-definite if and only if all of its 
eigenvalues are nonnegative.} square-root of 
\beqx
\Fb=\Gammab_E^{1/2}\Hb^+\Gammab_y\Hb^+\Gammab_E^{1/2},
\eeqx
where $^+$ denotes the generalized inverse~\cite[Chap. 5]{Golub96}. Indeed, matrices \Fb and \Sb respectively correspond to kernels $F$ and $f$ introduced in the proof of Property~\ref{Mercer} (see Appendix~\ref{proof}). Whereas $\Gammab_y$ quadratically depends on $\rhob$, $\Sb$ exhibits a linear dependency with respect to $\rho$, paving the way to an identifiability analysis \textit{via} a standard eigenvalue decomposition.

\subsection{Numerical estimation strategy}
\label{sec:numtrans_ES}

The reconstruction principle from the second-order data statistics amounts to finding $\rhob$ that makes the mean vector $\mub_z$ and the covariance matrix $\Gammab_z$ in \eqref{mom_vect} best match with the empirical quantities $\hat{\mub}_z$ and $\hat{\Gammab}_z$ defined by \eqref{eq:empirical}. 
Given the previous subsection, a simple idea to recover the identifiable components of $\wt\rhob$ would be to compute an approximation $\wh\Fb$ of matrix \Fb from the empirical data statistics:
$$
\wh\Fb=\Gammab_E^{1/2}\Hb^+\wh\Gammab_y\Hb^+\Gammab_E^{1/2},
$$
where $\wh\Gammab_y=\wh\Gammab_z-\Gammab_\varepsilon$, with a view to extract a positive semi-definite square-root matrix $\wh\Sb$. However, neither $\wh\Gammab_y$ nor $\wh\Fb$ are guaranteed to be positive semi-definite, so the existence of $\wh\Sb$ is not granted.

A preferable procedure consists in introducing an appropriate dissimilarity measure between the empirical and the theoretical second-order statistics of the data, and to minimize the dissimilarity to obtain an estimated sample $\wh\rhob$. One possible choice of dissimilarity measure is the Kullback-Leibler divergence (KLD) $D(\rhob)=D_{\text{KL}}(\Nc(\wh\mub_z,\wh\Gammab_z)\|\Nc(\mub_z,\Gammab_z))$, where $\Nc(\mub,\Gammab)$ is the normal distribution of mean $\mub$ and covariance $\Gammab$. According to~\cite[\S\,9.1]{Kullback59}, an explicit expression of $D(\rhob)$ is:
\bal
D(\rhob)=\;&\fracp12\tr \bigpth{\Gammab_z\M\wh\Gammab_z} + \fracp12(\mub_z-\wh\mub_z)\T\Gammab_z\M(\mub_z-\wh\mub_z)\notag\\
&+\fracp12\log\frac{\bars{\Gammab_z}}{\stdbars{\wh\Gammab_z}}-\fracp{N}2
\label{eq:DKL}
\eal
where $\bars{\cdot}$ and $\tr(\cdot)$ are the determinant and the trace of a square matrix, respectively. Let us mention that $D$ is proportional to the log-likelihood of the data under the assumption that the latter follow the normal distribution $\Nc(\mub_z,\Gammab_z)$ \cite{Miller87}. 
However, the minimizer of $D$ is an unregularized solution, which is unstable with respect to the random fluctuations in the dataset. Therefore, a penalization term must be added to $D$. In the sequel, we choose a quadratic penalization term to stabilize the solution, so that the SR effect remains purely driven by the data term. The criterion to minimize is then
\beq
\label{RegLik}
J(\rhob) = D(\rhob) + \frac{\beta}{2} \norm{\rhob}_2^2,
\eeq
with $\beta\geq 0$ and $\norm{\cdot}_2$ is the usual Euclidian norm. From a computational perspective, a closed-form minimizer cannot 
be found, so the minimization problem must be solved iteratively. Indeed, it is a so-called \textit{structured covariance} type problem, for which the Expectation-Maximization (EM) algorithm can be implemeted~\cite{Dempster77,Miller87,Lanterman00}. However, our tests indicate that the EM algorithm converges very slowly in the speckle SIM context. For this reason, we rather rely on a nonlinear conjugate gradient method, which turns out to produce more efficient iterations. It relies on the expression of the gradient of the penalized KLD \eqref{RegLik} with respect to $\rhob$ (see Appendix~\ref{Grad} for a derivation):
\bal
\label{gradDirectKL2}
&\nabla J(\rhob) = \nonumber\\
& -\left(\left[\Omegab\T\big( \Deltab_\Gamma+ \deltab_\mu\deltab_\mu\T\big)\Omegab\right] \circ\Gammab_E \right)\rhob - E_0 \Omegab\T\deltab_\mu + \beta\rhob,
\eal
displayed as a column vector, with $\Omegab=\Gammab_z\M\Hb$, $\deltab_\mu=\wh\mub_z - \mub_z$, $\Deltab_\Gamma= \wh\Gammab_z-\Gammab_z$, and $\circ$ stands for the Hadamard (component-wise) product.
Let us stress that each computation of the gradient needs the construction and the inversion of an $N\times N$ matrix (for an $N$-pixel size problem), which represents a prohibitive computing cost for realistic imaging problems. The design of less costly iterations for large-size problems is out of the scope of the present paper, but we are currently working on this crucial issue.

\subsection{Numerical illustration for 2D speckle SIM}
\label{sec:numtrans_NI}

\begin{figure*}[t]
 \centering
 \begin{tabular}{c@{\kern.2cm}c@{\kern.2cm}c}
  \includegraphics[height=5.4cm]{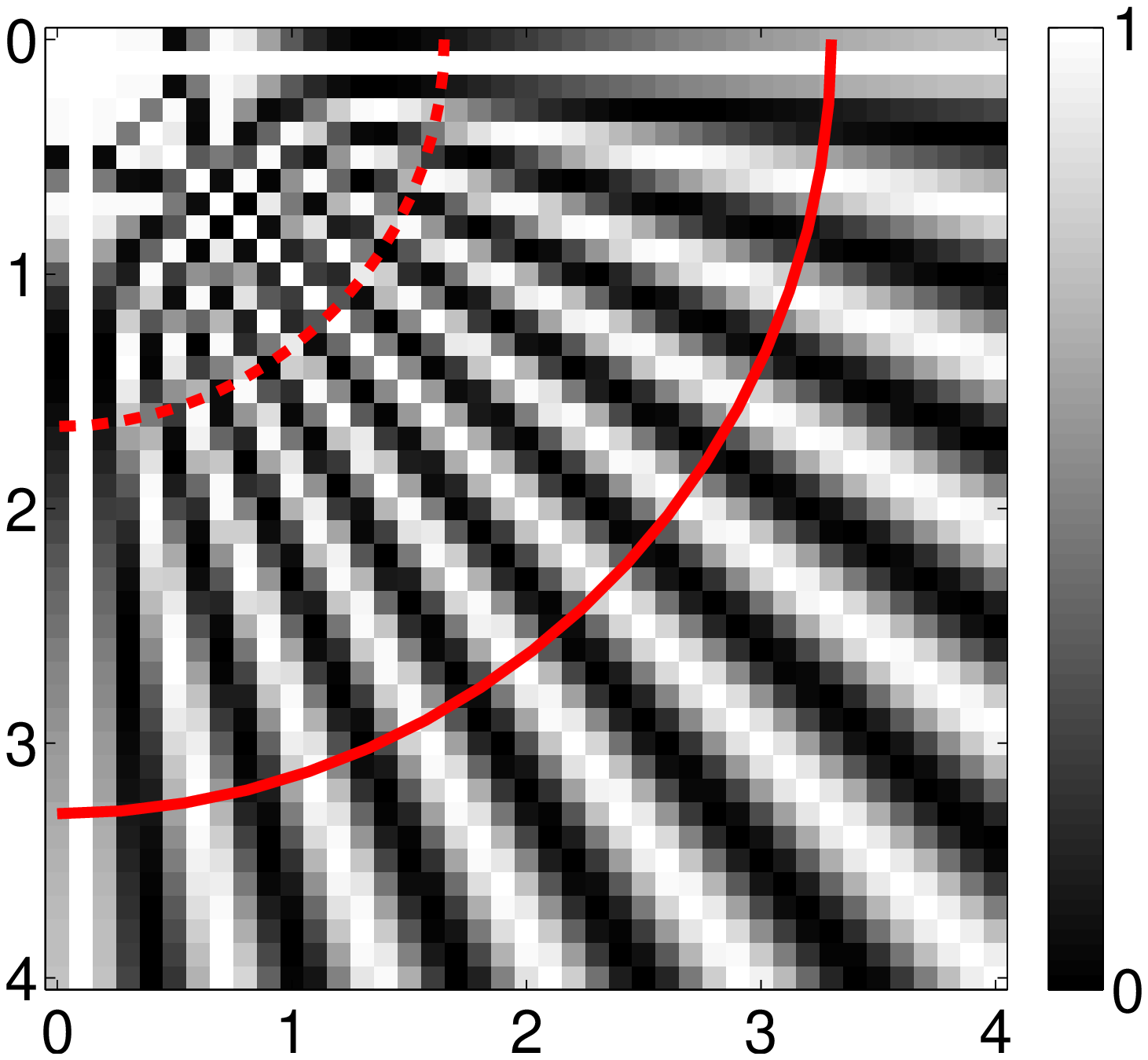}
 &
  \includegraphics[height=5.4cm]{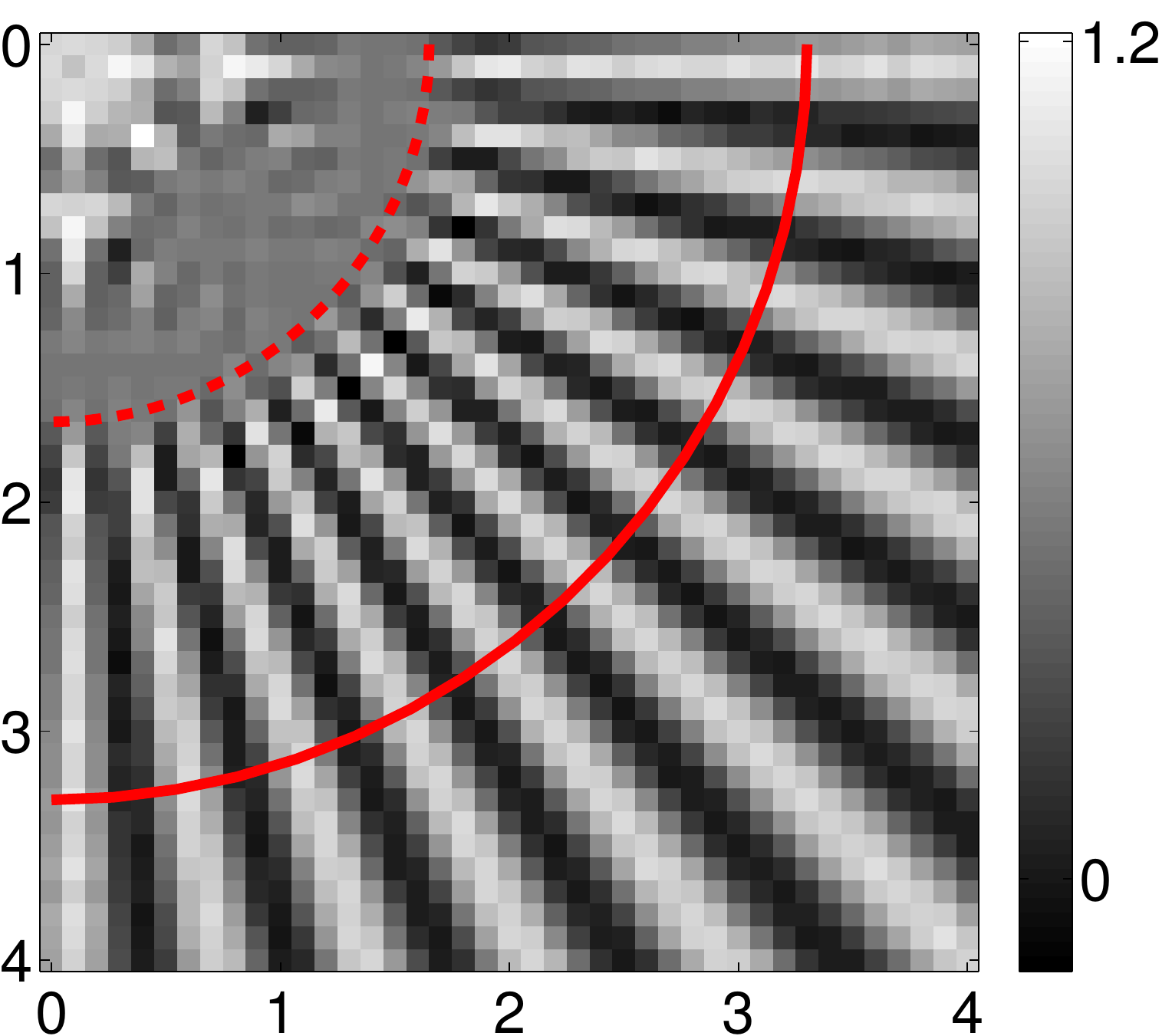}
 &
  \includegraphics[height=5.4cm]{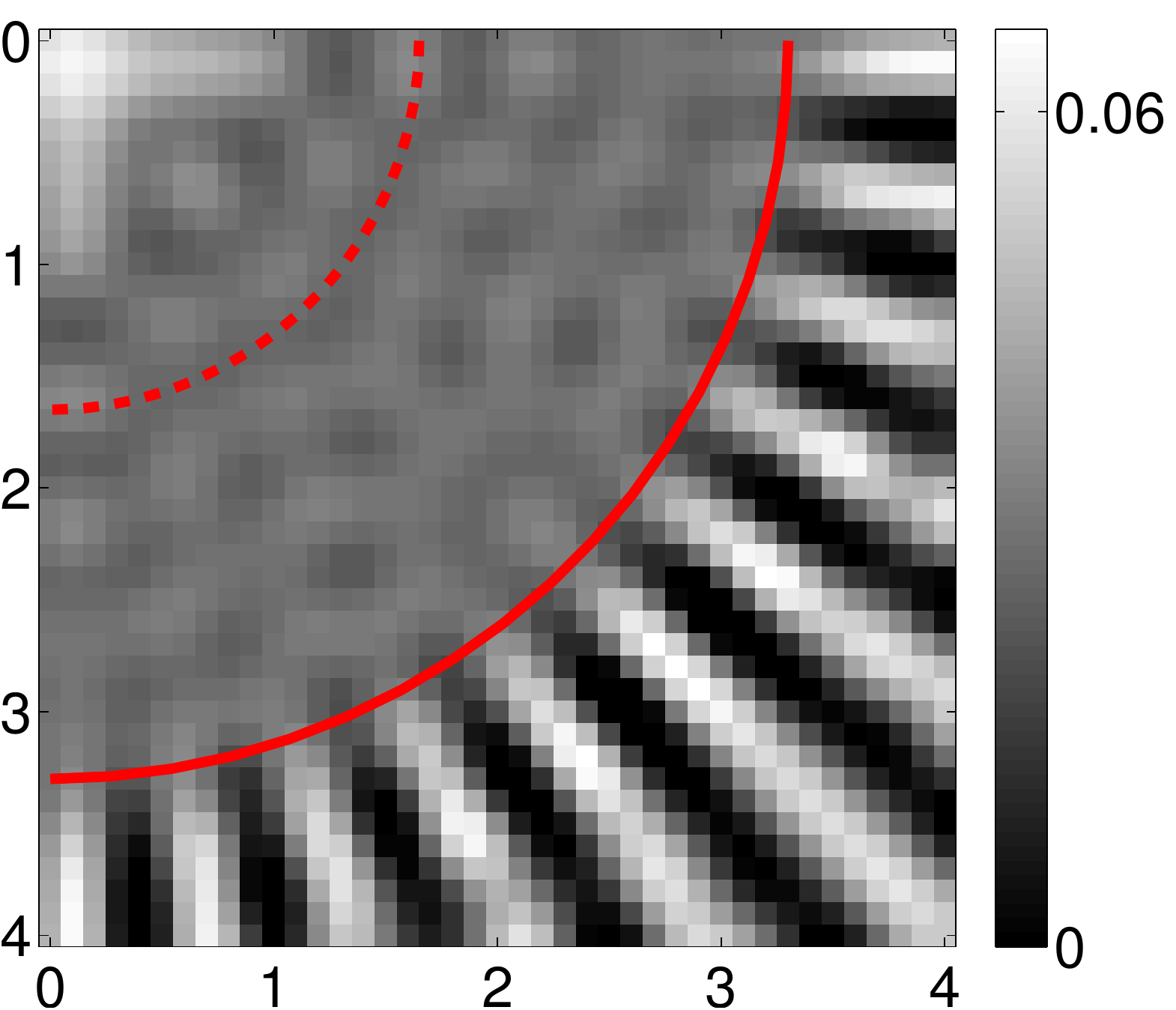}\\
 \small(a)&\small(b)&\small(c)\\[.5em]    
  \includegraphics[height=5.4cm]{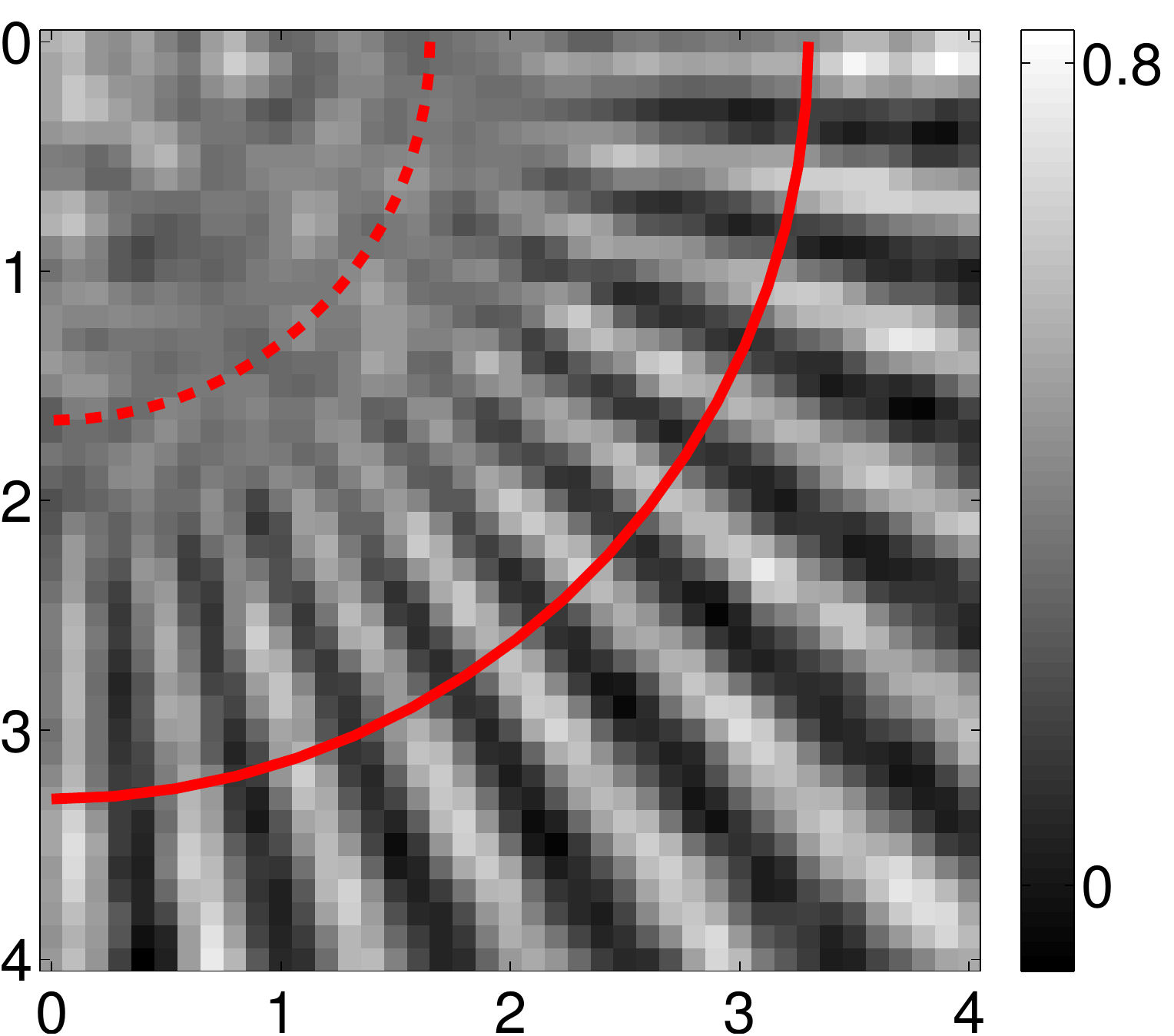}
  &
   \includegraphics[height=5.4cm]{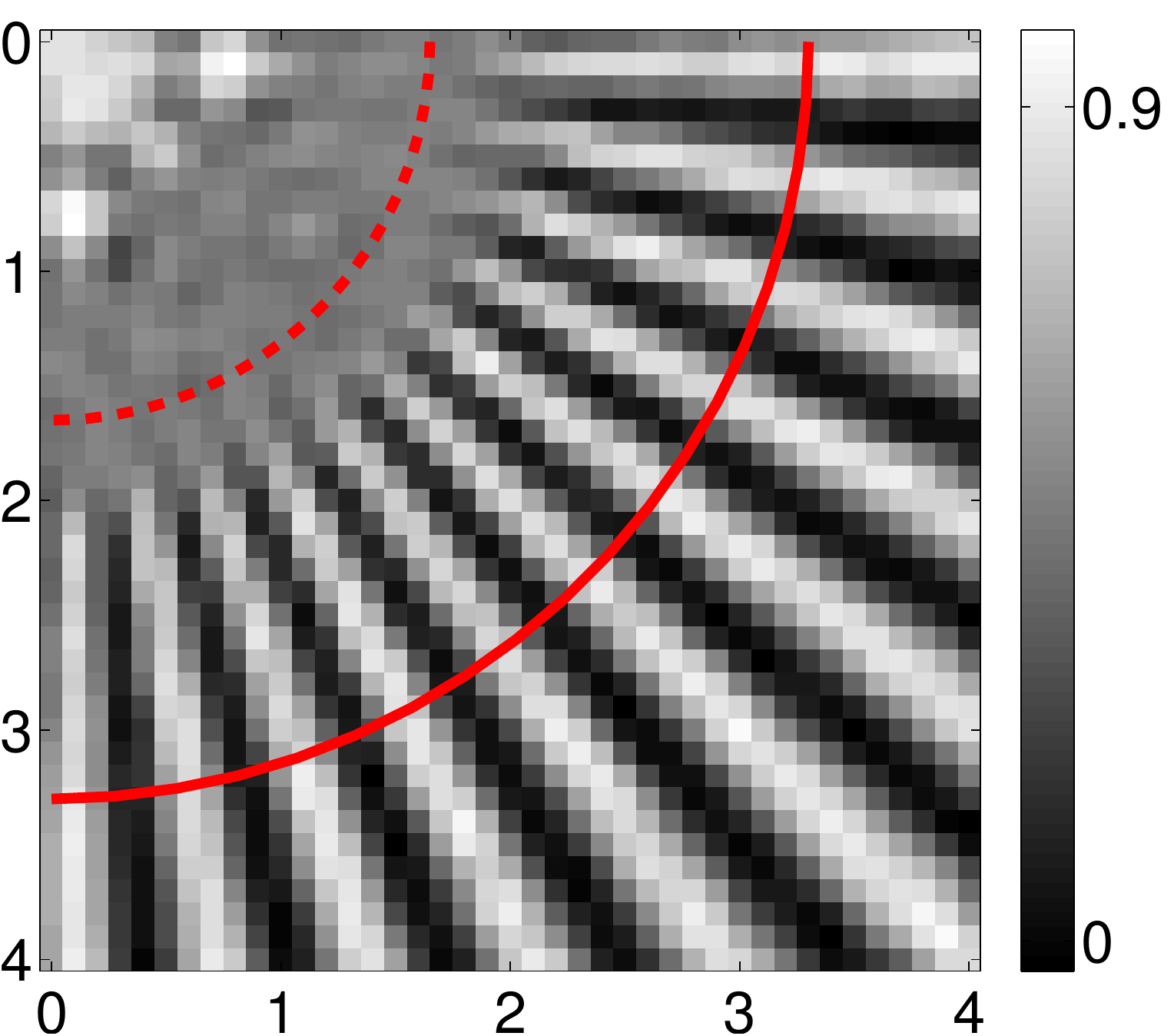}
  &
   \includegraphics[height=5.4cm]{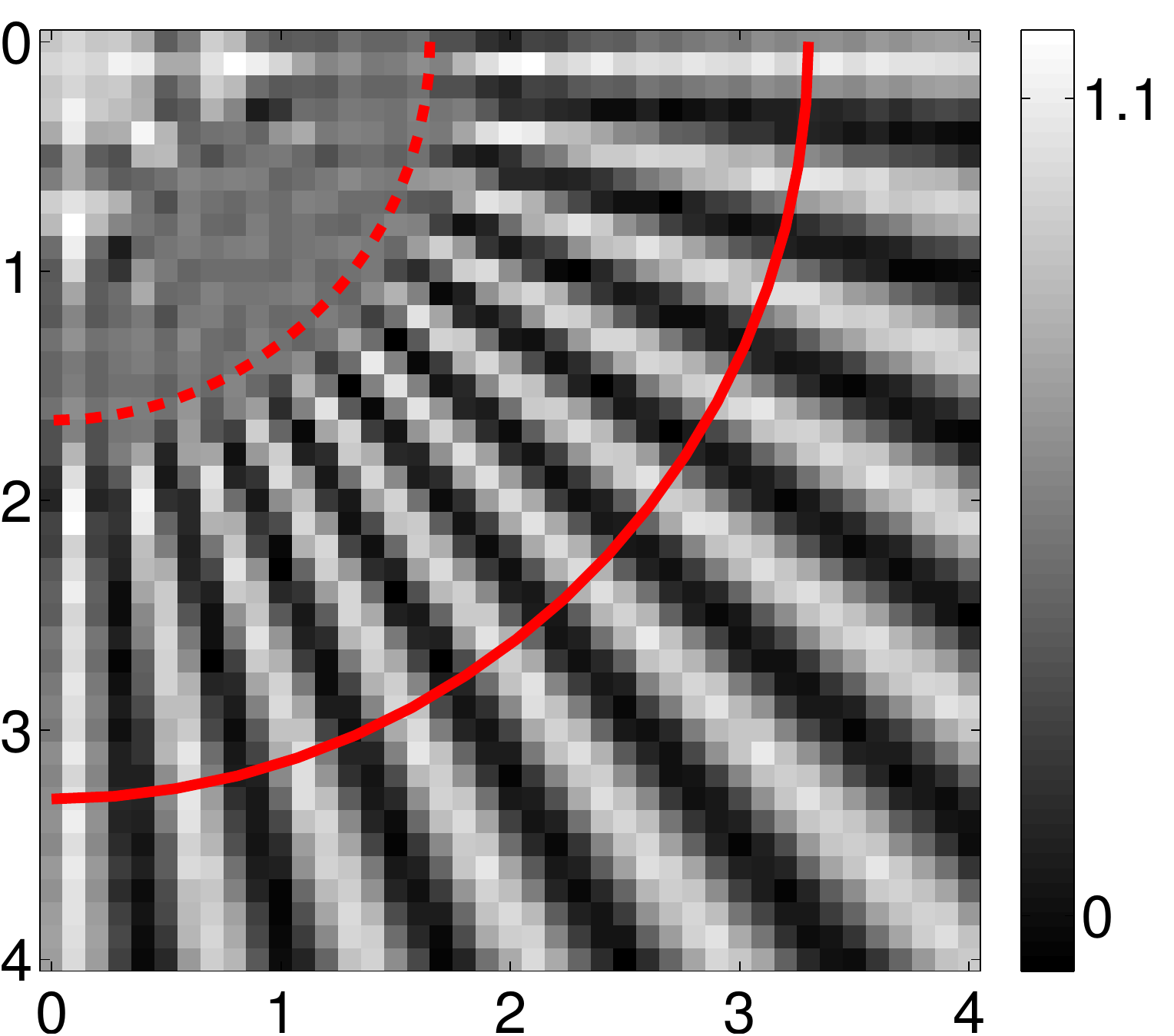}
  \\
  \small(d)&\small(e)&\small(f)
\end{tabular}
 \caption{%
  (a) Lower quarter of the (80$\times$80
  pixels) ground-truth fluorescence pattern considered in~\cite{Mudry12}.
  (b) Filtered ground-truth retaining only the
  spatial frequency lower than twice the OTF limit. (c) The deconvolution of the wide-field (constant illumination)
  microscope acquisition.
   (d,e) Estimator of $\rhob$ obtained from the
   minimization of the penalized KLD \eqref{RegLik} with $M=100$
   (d) and $M=1000$ (e) 
  speckle patterns; the regularization parameter is set to
  $\beta=\beta_0/M$ with $\beta_0=100$. (f) Estimator of 
 $\rhob$ obtained from the minimization the KLD \eqref{eq:DKL} with
 the asymptotic  statistics $\wh\mub = \mub^\star$ and $\wh\Gammab =
  \Gammab^\star$. 
  The distance units along the horizontal and 
  vertical axes are given in wavelength $\lambda$. The image sampling step
  for all simulations is set to $\lambda$/20. The dashed (resp. solid) lines corresponds to the spatial 
  frequencies transmitted by the OTF support (resp. twice the OTF
  limit).}
 \label{fig:fig1}
\end{figure*}

Numerical simulations are now considered to support that a significant SR effect can be obtained in speckle fluorescence SIM, even with a moderately large number of illumination patterns $\Eb_m$. The ground truth $\rhob^\star$ consists in the 2D 'star-like' fluorescence pattern depicted in Fig.~\ref{fig:fig1}(a). The convolution matrix $\Hb$ modeling the microscope is built from the discretized OTF associated with a circular aperture~\cite[Eq. (6-32)]{Goodman96}; the numerical aperture NA is set to 1.49 and the emission/excitation wavelength $\lambda$ is arbitrary set to 1. For this configuration, the resolution limit of standard wide-field imaging is clearly visible in Fig.~\ref{fig:fig1}(c). 
According to~\eqref{eq:observation3}, a set of $M\in\{100,1000\}$ speckle patterns are simulated to produce $M$ low-resolution microscope images $\{\zb_m\}_{m=1}^M$. The covariance matrix $\Gammab_E$ is set to $E_0^2\,\Hb$ (we assume $\gamma_E = E_0^2\,h$) and each acquisition $\zb_m $ is corrupted with an independent and identically distributed Gaussian noise such that the signal-to-noise ratio in each frame is set to 40 dB. From the dataset $\{\zb_m\}_{m=1}^M$, the statistics $\wh\mub_z$ and $\wh\Gammab_z$ \eqref{eq:empirical} are built.
The case of an infinite illumination number ($M=\infty$) is also adressed by considering the expected (\ie asymptotical) statistics $\wh\mub_z=\mub_z^\star$ and $\wh\Gammab_z=\Gammab_z^\star$, where $\mub_z^\star$ and $\Gammab_z^\star$ are obtained from \eqref{mom_vect} by setting $\rhob=\rhob^\star$. 
In all cases, we proceed to the iterative minimization of the penalized KLD \eqref{RegLik} to estimate the sample, using the deconvolved wide-field image of Fig.~\ref{fig:fig1}(c) as an initial point.
%
For the (noise-free) asymptotic statistics, the regularization parameter is set to $\beta=0$ and, as expected, the reconstruction exhibits the doubled resolution predicted by Prop.~\ref{Mercer}, see Fig.~\ref{fig:fig1}(f) compared to Fig.~\ref{fig:fig1}(b,c). With 100 and 1000 illuminations, the SR factor is lower, but the reconstructions shown on Figs.~\ref{fig:fig1}(d,e) are still much more resolved than the wide-field image of Fig.~\ref{fig:fig1}(c). Moreover, the SR factor progressively grows with the illumination number $M$, the result at $M=1000$ being close to the asymptotic regime.
%

\section{Conclusion and perspectives}
\label{sec:appli}

We have mathematically demonstrated that the mean and the \addJIF{covariance} function of low resolution images obtained with unknown, random illuminations permit to recover a super-resolved image of the sample, provided the first two statistical moments of the illuminations are fully characterized. Since this condition is expected to be less stringent to meet than the knowledge of each illumination, we believe that this result can be interesting in many practical situations.

In fluorescence microscopy, if the speckle is generated through the same objective as the one used to collect the light, its \addJIF{covariance function} is almost identical to the microscope PSF and Proposition~\ref{Mercer} is also expected to apply. We believe that this is a particularly important result. Indeed, it shows that speckle microscopy has the potential to generate a super-resolved image corresponding to the PSF $h_{\text{ext}}=|h|^2$. In other words, the SR would be equivalent to that of a perfect confocal microscope with infinitively small pinhole \cite{Kimura90}, but it would be obtained with no transverse scanning and with no loss of photons.

\addMAD{For coherent imaging system}, the consequence could be even more
spectacular. In holographic systems, the Fourier support of the PSF
$h$ is generally a cap of sphere. As a result, the three-dimensional
information on the sample is lost if only one illumination is
used. This is clearly observed in \addMAD{tomographic diffraction microscopy}
where the reconstruction of a target from its unique 2D hologram obtained under a monochromatic plane wave illumination is significantly deteriorated along one axis~\cite{Simon08}. On the other hand, by processing 2D images obtained under different speckle illuminations, one should be able to reconstruct the target in 
three dimensions with a PSF comparable to that obtained in tomographic diffraction imaging~\cite{Haeberle10}, but without the difficulty and slowness of controlling the angles of illuminations.

In photoacoustic imaging using speckle illuminations, the autocorrelation length of the random optical intensity can drop 
to a few hundreds of nanometers while the acoustic PSF $h$ has a typical width of tens of microns. Hence, from the acoustic point of view, the illumination can be seen as an uncorrelated random process \cite{Gateau13}, which is the case studied in Subsection \ref{ssec:uncorrel}. Using optical speckle in a photoacoustic experiment would allow to retrieve the square of the optical
absorption density with a resolution corresponding to the PSF $|h|^2$. 

Finally, it is important to stress the limits of the present analysis.
First, the case of complex-valued samples (\ie with both dielectric and absorptive components) remains to be investigated, since it could have important implications in electromagnetic tomography. The present study can be easily adapted to the case of pure absorptive (imaginary) samples, but an extension to the more general case is not so direct.
Second, our theoretical results are of asymptotic nature, in that they only predict the SR capacity of the imagers with an arbitrarily large number of illuminations. In particular, Proposition~\ref{Mercer} does not provide the \textit{sensitivity} of the retrievable sample frequency components. The simulation results shown in Section~\ref{sec:numtrans_NI} nonetheless suggest that these frequency components can be retrieved with only a few hundreds of illuminations.
Third, our results do not take into account the potential impact of advanced regularization in the inversion schemes (for instance, exploiting a sparsity prior~\cite{Min13} could yield an additional increase of resolution). 
Fourth, there exist many imaging configurations where the second-order
statistics do not entirely characterize the probability distribution
of the data (\eg when the speckle illuminations are positive
intensities). \addJIU{In such cases, our identifiability results only 
provide a lower bound on the super-resolution factor that could be 
reached from the complete data statistics, since}
accounting more precisely for the speckle statistics could still 
ameliorate the resolution. According to Remark~\ref{rmk}, \addJIF{no amelioration can be expected} when 
the support of the speckle \addJIF{covariance} identifies with that of
the PSF, but the question remains open in other cases. For instance, 
for uncorrelated speckles, one can write the following extension of 
Eq.~\eqref{eq:squared_conv},
\beq
\Cum_y^n(\rb\ldotsv\rb) = \Cum_E^n(\zerob\ldotsv\zerob)\,(\rho^n\otimes \bars{h}^n)(\rb),
\label{eq:sofi_like}
\eeq
where $\Cum^n$ denotes the $n$th circular cumulant of a given random process~\cite{Comon10}. Equation \eqref{eq:sofi_like} indicates that data higher-order statistics yield additional information on higher spatial frequencies of the sample. Such a property is reminiscent of the principle of SOFI~\cite{Dertinger09}.
%
The computational issue also remains broadly open, both in terms of
memory requirements (to store the empirical data covariance matrix)
and of computing time. The iterative scheme proposed in
Subsection~\ref{sec:numtrans_NI} is clearly limited to small-sized
images. A challenge will be to accelerate the reconstruction process 
while preserving the SR capacity of speckle-based imaging, as 
characterized in this paper.

\addMAU{\indent%
As a final remark, let us stress that controlled and random
illuminations lead in our opinion to distinct  ``resolution \textit{vs.} cost 
\textit{vs.} versatility'' trade-offs for the setup. In particular, 
when an accurate control of the illumination can be obtained 
within the sample volume, random illuminations may not 
be the best option to maximize the resolution for a given photon budget. 
On the contrary, random illuminations should achieve 
a better trade-off when the illumination cannot be controlled, 
or if one aims at designing \textit{versatile} and \textit{cheap} setups. 
The cautious evaluation of these trade-offs is a clear perspective of
this work.  
}

\section*{acknowledgments}
{The authors are grateful to the Associate Editor and to the anonymous reviewers for their 
valuable comments that helped in improving the quality of this article.}



\appendices
\section{Case of Poisson data}
\label{Poisson}

\add{%
For an incoherent imaging setup (\textit{e.g.,} optical fluorescence
microscopy), the intensity measurement relies on counting 
discrete particles and the model \eqref{eq:observation} 
can be replaced by 
\beq
 z_m(\rb) = p_m(\rb) + \varepsilon_m(\rb), \quad \rb\in\eZ^d
 \label{eq:observation2}
\eeq
where $p_m(\rb)$ is a Poisson random variable with mean 
$$
\Esp{p_{m}(\rb) \I E_m} = \int y_m(\zetab)\,\Pi(\zetab - \rb) \dd \zetab,
\quad \rb\in\eZ^d,
$$ 
where $\Pi$ is the indicator function of a centered detector pixel.
Assuming further that the pixel size is ``small'' with respect to 
the spatial variation in $y_m$, the expected counting rate for all 
$\rb \in \eZ^d$ is approximated by
$$
\Esp{p_{m}(\rb) \I E_m} =  a \, y_m(\rb) 
$$ 
where $a$ is the area of a single detector pixel.
We also assume that the Poisson outcomes are jointly 
statistically independent. 
The expression 
for $\mu_z(\rb)=\Esp{z_m(\rb)}=\Esp{p_m(\rb)}$ then reads, according to the law of iterated expectations:
$$
\Esp{p_m(\rb)}=\Esp{\Esp{p_m(\rb)\I E_m}} = a \, \Esp{y_m(\rb)}
$$
where $\Esp{y_m(\rb)}=E_0 (h\otimes\rho)(\rb)$.
%
Concerning the data \addJIF{covariance} function, we have for $\rb$, $\rb'\in\eZ^d$:
\balx
\gamma_z(\rb,\rb') & =\gamma_p(\rb,\rb')+\gamma_\varepsilon(\rb-\rb')
\ealx
with:
\beq
\gamma_p(\rb,\rb')
= \Esp{p_m(\rb) {p}_m(\rb')} - \Esp{p_m(\rb)}\,\Esp{{p}_m(\rb')}.
\eeq
According to the law of iterated expectations,
$$
\Esp{p_m(\rb){p}_m(\rb')}=\Esp{\Esp{p_m(\rb){p}_m(\rb')\I E_m}}.
$$
For $\rb'\neq\rb$, since $p_m(\rb)$ and $p_m(\rb')$ are decorrelated Poisson variables given $E_m$, we get
$$
\Esp{p_m(\rb){p}_m(\rb')}= a^2\, \Esp{y_m(\rb){y}_m(\rb')}
$$
while for $\rb'=\rb$,
$$
\Esp{p_m(\rb)^2} = a^2 \Esp{y_m(\rb)^2} + a \Esp{y_m(\rb)},
$$
since a Poisson variable is of equal mean and variance.
Therefore, \eqref{eq:correlation_z} must be replaced by
\beq
\gamma_z (\rb,\rb') = a^2\, \gamma_y(\rb,\rb') + \mu_z(\rb) \delta_\KD(\rb-\rb')+\gamma_\varepsilon(\rb-\rb')
\label{eq:correlation_z2}
\eeq
where $\gamma_y$ is given in \eqref{eq:correlation_y} and 
$\delta_\KD(\rb) = 1$ if $\rb=\zerob$ and zero otherwise.
}
\add{%
%
The data \addJIF{covariance} hence only differs from \eqref{eq:correlation_y} when $\rb=\rb'$. In this case, we note that the  additional term $\mu_z(\rb)$ is proportional  to $E_0$, whereas the variance $\gamma_y(\rb,\rb)$ varies as $\gamma_E(\rb)$,  which is usually proportional to $E_0^2$ for intensity speckle patterns~\cite{Goodman07}. Therefore, accounting for the Poisson  
statistics of the data in may be useful in the low counting-rate regime only.
}

\section{Proof of Property \ref{Mercer}}
\label{proof}

Let $q$ denote the impulse response of the filter defined in the Fourier domain by
$$
\wt q(\ub)=\wt\gamma_E^{1/2}(\ub)\quad\SI\ub\in \Dspec,\quad 0~\SINON.
$$
Akin to $\gamma_E$, $q$ is positive semi-definite, and hence it is a Hermitian symmetric function.
We have then $\wt\gamma_E=\wt q^2$, and hence
\beq
\gamma_E=q\otimes q.
\label{eq:qq}
\eeq
Let us also define the following kernels:
\bal
\label{eq:f}
f(\rb,\rb')&=\int q(\rb-\xb)\overline{q}(\rb'-\xb)\rho(\xb)\dd\xb,\\
F(\rb,\rb')&=\int f(\rb,\rb'') \overline{f}(\rb',\rb'')\dd\rb''.
\label{eq:F}
\eal
and the induced integral operators $K_f$ and $K_F$:
\balx
K_f\phi(\rb)&=\int f(\rb,\rb')\phi(\rb')\dd\rb',\\
K_F\phi(\rb)&=\int F(\rb,\rb')\phi(\rb')\dd\rb'.
\ealx

According to the Cauchy-Schwarz inequality, 
$$
\bars{f(\rb,\rb')}^2\leq\int \bars{q(\rb-\xb)}^2\rho(\xb)\dd\xb\int \bars{q(\rb'-\xb)}^2\rho(\xb)\dd\xb.
$$
As a consequence,
$$
\iint \bars{f(\rb,\rb')}^2\dd\rb\dd\rb'\leq\pth{\int \bars{q(\rb)}^2\dd\rb\int \rho(\xb)\dd\xb}^2,
$$
where $\rho$ is integrable, according to assumption (\emph{ii}), 
and
$$
\int \bars{q(\rb)}^2\dd\rb=\int\wt\gamma_E(\ub)\dd\ub=\gamma_E(0) < \infty.
$$
Therefore, we have
$$
\iint \bars{f(\rb,\rb')}^2\dd\rb\dd\rb'<\infty,
$$
\ie $f\in L^2(\eR^d\times\eR^d;\eC)$, and consequently, $K_f$ is a Hilbert-Schmidt integral operator~\cite[Proposition 3.4.16]{Pedersen95}.
On the other hand, the integral operator $K_F$ is the square of $K_f$, in the sense that $K_F\phi=K_f K_f\phi$ for any $\phi$. Thus,
$K_F$ is also a Hilbert-Schmidt operator.

Now let us go to the heart of the proof, which is threefold.
The first step allows us to show that kernel $F$ is uniquely defined from $\gamma_y$. In a second step, we establish
that $f$ is uniquely defined from $F$ given \eqref{eq:F}.
At this point, we conclude that the knowledge of $\gamma_y$ implies that of $f$, which is a linear functional of $\rho$ (whereas the dependency of $\gamma_y$ in $\rho$ is quadratic). The last step consists in a Fourier analysis of $f$, in order to determine which spectral components of $\rho$ are identifiable from the knowledge of $f$.
\medskip

\noindent\textit{Step 1)}
Given \eqref{eq:f} and \eqref{eq:qq}, we have the following alternate expression for \eqref{eq:F}:
\bal
&F(\rb,\rb')=
\FINAL{\notag\\&}
\iint\rho(\xb)\rho(\xb')\,q(\rb-\xb)\overline{q}(\rb'-\xb')\,
\gamma_E(\xb-\xb') \dd\xb \dd\xb'.
 \label{eq:Fbis}
\eal
Comparing the latter expression to \eqref{eq:correlation_y}, it is clear that $F=\gamma_y$ in the case $q=h$, \ie when the speckle \addJIF{covariance} is $h\otimes h$. More generally, 
using a double Fourier transform on \eqref{eq:Fbis}, in the same way as we obtained \eqref{eq:tgammay} from \eqref{eq:correlation_y}, we get
\bal
\wt F(\ub,\ub') &= \wt{q}(\ub) \wt{q}(-\ub')\wt{g}(\ub,\ub')\notag\\
&=\frac{\wt{q}(\ub) \wt{q}(-\ub')}{%
\addMAD{\wt{h}(\ub) \wt{h}^*(-\ub')}
}\wt{\gamma}_y(\ub,\ub')&&\SI\ub,\ub'\in\Dspec,
\label{eq:tildeF1}
\\
&=0&&\SINON.
\label{eq:tildeF2}
\eal
Let us remark that $\wt{h}(\ub)\neq0$ if $\ub\in\Dspec$ because we have assumed $\Dspec\subseteq\Dpsf$.
\medskip

\noindent\textit{Step 2)}
Kernel $f$ is obviously symmetric. 
Moreover, it is positive semi-definite, since for any square integrable function $\phi$,
$$
\iint f(\rb,\rb')\phi(\rb) \overline{\phi}(\rb')\dd\rb\dd\rb'=\int |q\otimes \phi|^2(\xb)\rho(\xb)\dd\xb\geq0.
$$
%
It is easy to check that kernel $F$ is also 
positive semi-definite. Moreover $K_F$ is bounded, since it is a Hilbert-Schmidt operator.
Being bounded and positive semi-definite, $K_F$ admits a unique square root~\cite[Prop. 3.2.11]{Pedersen95}. In other words, $K_f$ is uniquely defined given $K_F$, and equivalently, given the kernel $F$, there exists a unique kernel $f$ that fulfills \eqref{eq:F}.

Finally, the knowledge of $\gamma_y$ uniquely determines $F$ through \eqref{eq:tildeF1}-\eqref{eq:tildeF2}, which in turn determines $f$.
\medskip

\noindent\textit{Step 3)}
In the Fourier domain, Eq. \eqref{eq:f} reads
\beq
\wt{f}(\ub,\ub')=\wt{q}(\ub)\, {\wt{q}}(-\ub') \, \wt{\rho} (\ub+\ub').
\label{eq:step3}
\eeq
\addJID{The latter identity shows that $\wt{\rho}(\ub'+\ub'')$ is identifiable for all $(\ub',\ub'')$ such that $\ub'$ and $-\ub''$ belong to \Dpsf. We thus conclude that the frequency components $\wt{\rho}(\ub)$ are identifiable from kernel $f$, and thus from the data \addJIF{covariance} $\gamma_y$, for all $\ub\in\Dspec\ominus\Dspec$.}

\section{Gradient of the Kullback-Leibler divergence}
\label{Grad}

We first note that \eqref{eq:DKL} also reads 
\beq
D(\rhob) = \fracp12\log \det{\Gammab_z} + \fracp1{2M}
\tr\pth{\Gammab_z\M \Vb \Vb^t} + C
\label{alvm}
\eeq
where $C$ is an irrelevant constant term, and
\begin{equation}
 \label{Ehat0}
 \Vb = (\vb_1 | \cdots | \vb_M) \quad \text{with} \quad \vb_m=\zb_m-\mub_z. 
\end{equation}
The following identities~\cite[Sec.\,2]{Petersen12} will be useful for the derivation of the gradient of \eqref{alvm}:
\bal
\nabla_{\theta} \log \det{\Ab} &= \tr\pth{\Ab\M (\nabla_{\theta} \Ab)} \notag\\
\nabla_{\theta} (\Ab\M) &=-\Ab\M (\nabla_{\theta} \Ab) \Ab\M \label{g}\\
\nabla_{\theta} (\Ab \Bb) &= (\nabla_{\theta}\Ab) \Bb + \Ab (\nabla_{\theta}\Bb)\notag\\
\nabla_{\theta} \tr(\Ab) &= \tr\pth{\nabla_{\theta} \Ab}\notag\\
\nabla_{\theta} (\Ab\T) &= (\nabla_{\theta} \Ab)\T\notag
\eal
where $\Ab$ and $\Bb$ are two matrices that depend on a real scalar parameter $\theta$. From these relations, we get
\bal
\partial_n D(\rhob) &= \fracp12\tr\pth{\Gammab_z\M (\partial_n \Gammab_z)}\notag\\
&+\fracp1{2M} \tr\pth{(\partial_n\Gammab_z\M) \Vb\Vb\T + \Gammab_z\M \partial_n(\Vb\Vb\T)}
\label{Dalvm}
\eal
where $\partial_n = \nabla_{\rho_n}$. The gradient of \eqref{alvm} is then defined by
\bal
\nabla D(\rhob) =&~ \vectb\acc{\partial_n D(\rhob)}
\label{eq:grad}
\eal
where $\vectb\acc{v_n}=\pth{v_1|\cdots|v_N}\T$.
According to \eqref{Dalvm} and \eqref{g}, the expressions of $\partial_n\Gammab_z$ and
$\partial_n(\Vb\Vb\T)$ are required. Let $\eb_n$ be the $n$th canonical vector, $\hb_n$ the $n$th column of 
$\Hb$ and $\unb=(1\cdots 1)\T$. We get from \eqref{mom_vect}
\bal
\partial_n\Gammab_z
&= \Hb\Rb\Gammab_E \eb_{n}\hb_n\T + (\Hb\Rb\Gammab_E \eb_{n}\hb_n\T)\T
\label{Dalvm_b}
\eal
and from \eqref{Ehat0}:
\bal
\partial_n(\Vb\Vb\T) 
&= - E_0 \pth{\Vb\unb\hb_n\T + (\Vb\unb\hb_n\T)\T}.
\label{Dalvm_a}
\eal
The derivative of the three terms in \eqref{Dalvm} can now be
obtained. On the one hand, elementary manipulations involving 
the trace operator allow to deduce
\bal
\tr\pth{\Gammab_z\M (\partial_n \Gammab_z)} &=2\,\eb_n\T\Wb\hb_n
\label{Dalvm_1a}
\eal
from \eqref{Dalvm_b}, with $\Wb = \Gammab_E\Rb\Hb\T\Gammab_z\M$. On the other hand, 
we have from \eqref{g} and \eqref{Dalvm_b}:
\bal
\tr\pth{(\partial_n\Gammab_z\M) \Vb\Vb\T} =&~-2\,\eb_n\T(\Wb\Vb\Vb\T\Gammab_z\M)\hb_n
\label{Dalvm_1c}
\eal
and from \eqref{Dalvm_a}:
\bal
\tr\pth{\Gammab_z\M \partial_n(\Vb\Vb\T)} =-2\, E_0 \,\hb_n\T \Gammab_z\M\Vb\unb.
\label{Dalvm_1d}
\eal
According to \eqref{Dalvm} and \eqref{eq:grad}, we need to vectorize the relations \eqref{Dalvm_1a},
\eqref{Dalvm_1c} and \eqref{Dalvm_1d} to obtain the full gradient of
\eqref{alvm}. In particular, according to the identity 
\beq
\label{a}
((\Ab\,\Diag(\rhob)\Bb^t) \circ \Ib) \unb = (\Ab \circ \Bb) \rhob,
\eeq
%
we deduce from \eqref{Dalvm_1a} that
\bal
\vectb \acc{\tr\pth{\Gammab_z\M (\partial_n \Gammab_z)}}
&=2\left( (\Wb\Hb) \circ \Ib \right) \unb \notag\\
&=2\big( (\Hb\T\Gammab_z\M\Hb) \circ \Gammab_E \big)\rhob. \label{Dalvm_1b}
\eal
Similarly, we obtain after a few manipulations
\beq
\frac1M \vectb \acc{\tr\pth{(\partial_n\Gammab_z\M)\Vb\Vb\T} } =-\fracp2M\big((\Omegab\T\Vb\Vb\T\Omegab) \circ \Gammab_E \big)\rhob
\eeq
and
\beq
\frac1M \vectb \acc{ \tr\pth{\Gammab_z\M \partial_n(\Vb\Vb\T)}} =-\frac2M 
E_0 \Omegab^t\Vb\unb 
\eeq
where $\Omegab=\Gammab_z\M \Hb$. As a result, the gradient of
\eqref{Dalvm} reads
\bal
\label{eq:grad1}
&\nabla D(\rhob) =\notag\\
&-\left( \pth{\Omegab\T \pth{\fracp1M\Vb\Vb\T-\Gammab_z}\Omegab} \circ \Gammab_E \right)\rhob -\fracp1M E_0 \Omegab^t\Vb\unb. 
\eal
Finally, the following relations hold:
\balx
 \Vb &= (\zb_1-\wh\mub|\cdots|\zb_M-\wh\mub) + \deltab_\mu \unb\T,\\
 \fracp1M \Vb\Vb\T &= \wh\Gammab + \deltab_\mu \deltab_\mu\T,
\ealx
%
which allow us to obtain the gradient expression \eqref{gradDirectKL2}, given that $\nabla J(\rhob) = \nabla D(\rhob) + \beta\rhob$. 


\bibliographystyle{ieeeji}
\bibliography{bibenabr,revuedef,BlindsimInfo}

\end{document}